\documentclass[a4paper,11pt]{article}
\pdfoutput=1 

\usepackage{jheppub} 

\usepackage[T1]{fontenc} 

\usepackage{pgfplots}
\usepackage{tikz}
\pgfplotsset{compat=1.14} 

\usepackage{amssymb,amsmath,epsfig,latexsym,booktabs,braket,
gensymb,amsbsy,stackrel,slashed}

\title{\boldmath Multi-particle finite-volume effects for hexagon tessellations}

\author[a]{Marius de Leeuw,}
\author[b]{Burkhard Eden,}
\author[b]{Dennis le Plat,}
\author[b]{Tim Meier,}
\author[c]{Alessandro Sfondrini}

\affiliation[a]{School of Mathematics \& Hamilton Mathematics Institute,\\
Trinity College Dublin,
20 Westland Row, Dublin 2,
Ireland}
\affiliation[b]{Institut f\"ur Physik, Humboldt-Universit\"at zu Berlin,\\
IRIS Geb\"aude, Zum Grossen Windkanal 6, 12489 Berlin, Germany}
\affiliation[c]{Institut f\"ur theoretische Physik, ETH Z\"urich\\
Wolfgang-Pauli-Strasse 27, 8093 Z\"urich, Switzerland}

\emailAdd{mdeleeuw@maths.tcd.ie}
\emailAdd{eden@math.hu-berlin.de}
\emailAdd{diplat@physik.hu-berlin.de}
\emailAdd{tmeier@physik.hu-berlin.de}
\emailAdd{sfondria@itp.phys.ethz.ch}

\abstract{Correlation functions of gauge-invariant composite operators in ${\cal N}=4$ super Yang-Mills theory can be computed by integrability using triangulations. The elementary tile in this process is the hexagon, which should be glued by appropriately inserting resolutions of the identity involving virtual (``mirror'') magnons. 
We consider this problem for five-point functions of protected operators. At one-loop in the 't Hooft coupling, it is necessary to glue three adjacent tiles which involves two virtual magnons scattering among each other.
We show that the result can be simplified by using an adapted mirror rotation and employing appropriate summation techniques. The mirror-particle contributions then yield hyperlogarithms of weight two.
Finally, we use these results to investigate braiding prescriptions introduced in earlier work on the problem.

}

\preprint{HU-EP-19/42}

\newcommand{\beq}{\begin{equation}}
\newcommand{\eeq}{\end{equation}}
\newcommand{\rar}{\, \rightarrow \,}
\newcommand{\bino}[2]{\left( \begin{array}{c} #1 \\ #2 \end{array} \right)}
\newcommand{\gfunc}[1]{\Gamma\left[#1\right]}
\newcommand{\cX}{{\cal X}}
\newcommand{\cY}{{\cal Y}}
\newcommand{\cZ}{{\cal Z}}
\newcommand{\mS}{\mathbb{S}}
\newcommand{\ra}{\rangle}
\newcommand{\la}{\langle}

\begin{document}

\maketitle

\section{Introduction}

The AdS/CFT conjecture connects ${\cal N} = 4$ super Yang-Mills theory in four dimensions to IIB string theory on AdS$_5\times$S$^5$, see \textit{e.g.}~\cite{Aharony:1999ti} for a review.
In particular, the spectrum of anomalous dimensions in field theory should correspond to the energy levels of the string. The BMN construction \cite{Berenstein:2002jq} gave the first example of a class composite of operators in field theory and the dual string states. Subsequently, the computation of the planar part of the one-loop anomalous dimensions of the BMN operators has been interpreted as the diagonalisation of an integrable spin-chain Hamiltonian \cite{Minahan:2002ve}, thereby linking the AdS/CFT correspondence to an integrable system.\footnote{See refs.~\cite{Arutyunov:2009ga,Beisert:2010jr} for reviews of the integrability approach to AdS/CFT.}
A deformation of the corresponding Bethe equations can incorporate the effects of planar higher-loop corrections in the field theory. This is achieved by using the Zhukowski variables $x(u)$ defined by
\beq
x + \frac{g^2}{2 \, x} \, = \, u \label{defXRoot}
\eeq
instead of the Bethe rapidity $u$ of the original spin chain. Since the Bethe ansatz involves the quantities $u^\pm \, = \, u \pm \frac{i}{2}$ one further introduces $x^\pm(u) \, = \, x(u^\pm)$. In the definition (\ref{defXRoot}), $g^2 \, = \, g^2_\mathrm{YM} N / (8 \pi^2)$ is the 't Hooft coupling involving $N$, the rank of the gauge group $SU(N)$.

All fields of the ${\cal N}=4$ SYM theory transform in the adjoint representation of $SU(N)$. Gauge-invariant local operators arise by taking traces of products of such field, perhaps including gauge-covariant derivatives. The deformed integrable model has been extended to the complete set of single-trace operators~\cite{Beisert:2005fw,Beisert:2005tm}, so when all constituent fields are united in one gauge group trace. When supplied with the ``dressing phase'' it apparently correctly captures the planar anomalous dimensions to any desired order in the coupling constant~\cite{Beisert:2006ez}, provided that an asymptotic regime is not left, \textit{i.e.}\ that the length of the spin chain (the number of fields the single-trace operator consists of) is greater than the loop order. Beyond this bound the group factors of Feynman graphs change invalidating the approach. Quite literally this is a ``finite size effect'' because connected Feynman graphs start to wrap around the operator at this order, as it is most clearly understood from the dual string-worldsheet picture~\cite{Ambjorn:2005wa}.
Indeed, viewing the dual string as a two dimensional field theory, these effects may be computed using the thermodynamic Bethe ansatz (TBA) \cite{Zamolodchikov:1989cf} for a ``mirror'' model~\cite{Arutyunov:2007tc}. As for relativistic models~\cite{Zamolodchikov:1989cf}, a double Wick rotation exchanges space and time. In the case at hand one obtains a mirror theory that is not identically equal to the original model. Nonetheless, the two theories are related by analytic continuation. In terms the the Bethe rapidities~$u$, the mirror transformation is%
\footnote{This transformation can also be realised as an imaginary-rapidity shift, much like in relativistic models, by introducing a suitable elliptic parametrisation of the rapidities, \textit{cf.}~\cite{Arutyunov:2009ga}.}
\beq
\gamma  : \qquad x^+ \, \rightarrow \, \frac{1}{x^+},\quad x^-\to x^-\,.
\eeq
As it is the case in relativistic models, this transformation is ``half'' of a crossing transformation~\cite{Janik:2006dc}. In this sense, we can define a ``$2\gamma$'' transformation which yields crossing (flipping the sign of energy and momentum), and by iteration a $3\gamma$ and $4\gamma$ transformation:
\begin{equation}
\begin{aligned}
2 \gamma : & \qquad x^+ \, \rightarrow \, \frac{1}{x^+},\quad &&
\qquad x^- \, \rightarrow \, \frac{1}{x^-}
\, , \\
3 \gamma : & \qquad x^+ \, \rightarrow \, x^+,\quad &&
\qquad x^- \, \rightarrow \, \frac{1}{x^-}
\, , \\
4 \gamma : & \qquad x^+ \, \rightarrow \, x^+,\quad &&
\qquad x^- \, \rightarrow \, x^-
\, .
\end{aligned}
\end{equation}
It should be stressed that this transformation entails an analytic continutation, and therefore a $4n\gamma$ transformation is the identity only on analytic functions of $x^\pm$,%
\footnote{More precisely, $4n\gamma$ is the identity on functions which are meromorphic on a suitably-defined rapidity torus~\cite{Arutyunov:2009ga}. This includes meromorphic functions of~$x^\pm$ and the square-root functions which appear in Beisert's S-matrix~\cite{Beisert:2005tm}. } while functions with cuts need to be analytically continued. Importantly, this is the case of the dressing phase~\cite{Beisert:2006ez}, see~\cite{Arutyunov:2009kf}.
The TBA for the AdS/CFT correspondence is fairly involved, as it requires considering all bound states of excitations of the mirror model~\cite{Arutyunov:2007tc}. The symmetry structure of the direct and mirror theory is based on the same $su(2|2)$ superalgebra, up to an analytic continuation. This algebra allow two types of short bound-state representations, the symmetric and the antisymmetric representation. As it turns out, the former corresponds to bound states of the direct theory, while the latter to bound states of the mirror theory (the ones important for the TBA construction). The scattering matrix for such bound states and its dressing phase have been studied in refs.~\cite{Arutyunov:2009mi,Arutyunov:2009kf}. 

For a long time, higher-point functions remained hard to compute through integrability. The introduction of the ``hexagon operator'' was a breakthrough for the three-point problem \cite{Basso:2015zoa}. The $\mathcal{N}=4$ SYM three-point problem is related to the splitting of a closed string into two, with each external closed-string state being a cyclic spin-chain state. Topologically, the worldsheet is a sphere with three punctures, which can be triangulated yielding two patches. Each patch is ``hexagonal'': three edges correspond to the triangulation cuts, and three to pieces of the closed-string states. The expectation value of such an object in presence of  excitations on the closed-string states is given by an integral form factor. One can also consider excitation on the other, ``mirror'' edges (those corresponding to the triangulation cuts). In that case, excitations have the mirror kinematics~\cite{Basso:2015zoa}. Indeed moving excitations from one edge to another amounts to analytically continuing it by $n\gamma$. In practice, to evaluate this form factor one has to compute an object proportional to Beisert's S-matrix for magnons in an $n\gamma$ kinematics. While it is an outstanding problem to formulate a mirror-TBA-like approach for three point functions, it is possible to account for the leading wrapping effects by introducing mirror excitations on the mirror edges, summing over all allowed flavours (including bound states) and integrating over their momenta, in a spirit similar to L\"uscher's formula~\cite{Luscher:1985dn,Luscher:1986pf}. It should be stressed that this approach is perturbative in the 't~Hooft coupling,%
\footnote{Much like it is the case for the spectrum, the large-volume expansion in the sense of L\"uscher becomes a perturbative expansions in the 't~Hooft coupling once the mirror kinematics is properly accounted for~\cite{Ambjorn:2005wa}.}
and becomes quite involved computationally beyond the first few orders~\cite{Eden:2015ija,Basso:2015eqa,Basso:2017muf}.

\begin{figure}[t]
\begin{center}
\includegraphics[width=\linewidth]{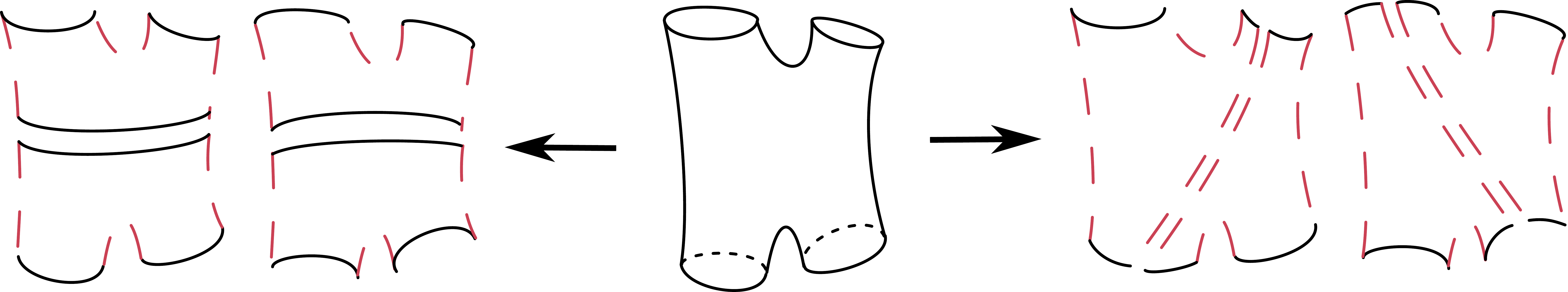}
\end{center}
\caption{The lowest-genus four-point function is topologically a sphere with four punctures corresponding to the closed-string states (middle). There are two way to tessellate this object in hexagonal patches. On the left, we cut the four-point function into two three-point functions, like in the operator-product expansion (OPE), and tessellate them with hexagons. This requires summing over all intermediate \emph{physical} states (solid lines). In the right-hand panel, the cut is only along mirror edges (dashed lines), so we only need to sum over all possible exchanges of mirror particles. The latter approach is often computationally more powerful, and it is our focus here.}
\label{fig:1}
\end{figure}

The construction of \cite{Basso:2015zoa} was extended to higher-point functions in \cite{Eden:2016xvg,Fleury:2016ykk}. This requires more general tessellations, see Figure \ref{fig:1}. 
Also in this case, it is necessary to introduce mirror particles and bound states, summing over all allowed flavours and integrating over the particle's momenta. One further complication is that four- and higher-point functions depend on conformal cross ratios as well as two more cross-ratios describing the R-symmetry configuration. This dependence should twist the sums over mirror particles, though how this should precisely happen is, at this stage, still a conjecture. Testing this conjecture further is one of our goals.   

\begin{figure}[t]
\begin{center}
\includegraphics[width=0.85 \linewidth]{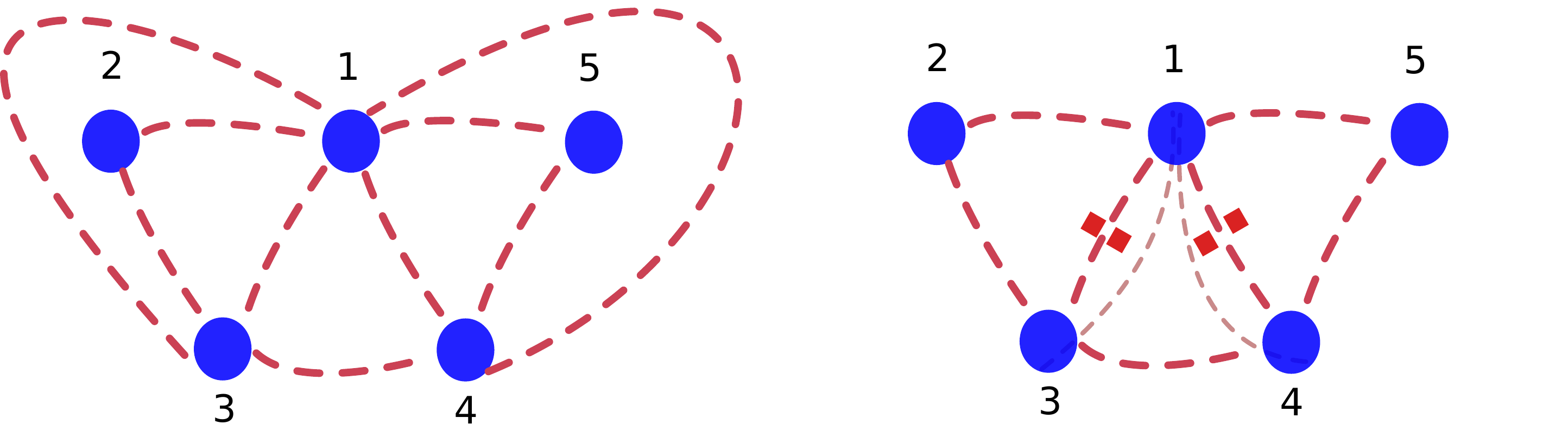}
\end{center}
\vskip -0.4 cm
\caption{We consider the tessellation of a five-point function involving five protected operators (blue dots). Left: we tessellate the figure by mirror cuts (dashed lines), which can be further decorated by inserting pairs of mirror magnons. Right: the object of our interest is when two pairs of virtual particles (red boxes) sit on edges $13$ and $14$l in this case, the hexagon $143$ contains two mirror particles, which gives rise to a non-trivial mirror-mirror scattering. Note that the mirror particles are arbitrary bound-states and have arbitrary momenta. As we will often do, we represented the outermost cuts ``behind'' the figure (using lighter lines).}
\label{fig:2}
\end{figure}

The simplest set-up to consider multi-particle contributions to finite-volume effects appears for five-point functions. For simplicity we may consider five-point functions of protected (\textit{i.e.}, 1/2-BPS) operators and focus on their cross-ratio dependence. In this case, we can have two pairs of mirror excitations, along two cuts shared by one of the patches, see Figure \ref{fig:2}.
The process has been evaluated in \cite{Fleury:2017eph} by matching a truncated residue calculation on an ansatz inspired from field-theory considerations~\cite{Drukker:2008pi}. We want to revisit such a computation. The motivation for our study is threefold: firstly, there are two distinct $n\gamma$ rotations of the process, which lead to widely different sums over residua. Do the two scenarios really agree, as they must? Secondly, we need to perform a double sum over the bound-state numbers, and a double integration over their momenta; can this be done in closed form, without resorting to any ansatz? Finally, in \cite{Fleury:2017eph} it was proposed that additional braiding factors $e^{\frac{i}{2} p}$, where $p$ denotes the momentum of the bound state particles, have to be introduced. This was a somewhat ad-hoc prescription, and it would be important to understand whether the choice of braiding adopted there is the only possible one and why.

This paper is organised as follows: in Section \ref{sec:1} we collect the many ``ingredients'' for the actual calculations. In Section \ref{sec:2} we put these together and set out on integrating/summing where we have methods to do so. We find weight two hyperlogarithms term-wise. The results suggest a function space on which we can fit the remaining sectors of the calculation. Section \ref{sec:3} starts with a review of the one-magnon process of \cite{Fleury:2016ykk} trying to draw conclusions about the so-called $J$-charge. Then we assemble the parts of our five-point two-magnon computation for one specific tessellation. The hexagon tiling idea obviously only makes sense when all results are triangulation independent, which for our choice still imposes invariance under cyclic permutations. This constraint is non-trivial and turns out to uniquely determine the result, although the $J$-charge and momentum dressing are not completely pinned down. The outcome does allow a prescription similar to that of \cite{Fleury:2017eph}, but not identical.

Four appendices collect simplifications of the bound-state $S$-matrix, the complete re-summation that we can achieve in case of the diagonal $\cX$ elements of this $S$-matrix, an alternative derivation of the complete result for diagonal $\cZ_{11}$ elements to the one presented in the main text, and finally a re-analysis of the structure constant computation in \cite{Basso:2015zoa} from the point of view of the exchange of bound states, to ensure that our new prescription is backward-compatible, so to say.

\section{Elements of the calculation}
\label{sec:1}

To begin with, we consider the expectation value of the hexagon $123$ and the hexagon $134$ in Figure \ref{fig:2} in the presence of one pair of virtual particles on the edge $13$. The virtual particles couple the two hexagons, yielding a non-trivial result. In absence of other (virtual or real) excitations, the remaining hexagons give trivial contributions. This particular computation was first considered in~\cite{Fleury:2016ykk}.

By conformal transformations we can always move the points into a plane at  $x_1  =  0,$ $x_2  =  1,$ $x_3 = \infty,$ and $ x_4 = ( 0,-\Im(1/z), \Re(1/z), 0 )$ for some complex parameter~$z$. The first hexagon now connects the points $0,1,\infty$ as in the defining three-point problem studied in \cite{Basso:2015zoa}; the configuration is ``canonical''. The second hexagon shares the edge between $0,\infty$ over which we glue, but its third point is parametrised by the variables $z,\bar z$.
For the two independent conformal cross-ratios on which the process will depend we find
\beq
\frac{x_{14}^2 x_{23}^2} {x_{12}^2 x_{34}^2} \, = \, z \bar z \, , \qquad \frac{x_{13}^2 x_{24}^2}{x_{12}^2 x_{34}^2} \, = \, (1-z)(1-\bar z) \, .
\eeq
In \cite{Fleury:2016ykk} it is suggested to obtain the non-standard coordinates $z,\bar z$ from the usual situation $0,1,\infty$ acting on the second hexagon with the operator
\beq
W(z,\bar z) \, = \, e^{-D \, \log|z|} \Big(\frac{z}{\bar z}\Big)^{L/2} \, , \qquad L \, = L^1_1 - L^2_2 \label{rotateHex}
\eeq
where $D$ is the dilatation generator. The idea is now to trade the representation of $D$ on the coordinates for its representation on spin-chain eigenstates
\beq
\frac{1}{2} (D-J) \, = \, E \, = i \tilde p \, = \, i \, u + \ldots
\eeq
where $\tilde p$ refers to the momentum in mirror kinematics, and the dots indicate terms of higher order in the coupling~$g^2$. We will discuss the effects of the charge $J$ when collecting results in Section \ref{sec:3}. The operator $W(z,\bar z)$ is applied to the bound states inserted on the common edge $13$. Since the generators employed are diagonal this introduces a multiplicative weight factor. The hexagons themselves can finally be evaluated as in the three-point problem.

To compute the full five-point functions we will need to consider virtual particles on other mirror edges. In particular, some of these will sit in one of the hexagons involving the operator at $x_5$. We used conformal symmetry to move $x_1,x_2,x_3$ and $x_4$ in a convenient configuration; the same cannot be done with $x_5$. We will however assume, following \cite{Fleury:2017eph}, that $x_5$ lies in the plane generated by the other four points. This is a restriction with respect to the most general five-point kinematics. The entire five-point function will then only rely on computing weights of the form $W(z_1,\bar z_1) \, W(z_2,\bar z_2)$, where $z_1$, $z_2$ are different cross ratios (related to points $1234$ and $1345$) but refer to the same plane.

Next, we have to sum over all relevant bound states. Following ref.~\cite{Basso:2015zoa}, we will sum over the mirror-theory bound states~\cite{Arutyunov:2007tc}, \textit{i.e.}\ the $sl(2)$-sector bound states. This ``antisymmetric representation'' at level (or length) $a$ has the basis
\beq
(\psi^1)^{a-k-1} (\psi^2)^k \phi^i \, , \qquad (\psi^1)^{a-k} (\psi^2)^k \, , \qquad (\psi^1)^{a-k-1} (\psi^2)^{k-1} \phi^1 \phi^2 \, . \label{sl2BoundBits}
\eeq
In the last formula, $\phi^i$ are a doublet of bosonic scalar constituents, while $\psi^\alpha$ are two-component spinors. Customarily, for the first class of states one separately considers $i \, = \, 1,2$.
Now,
\beq
W(z,\bar z) \, (\psi^1)^{a-k} (\psi^2)^k \, = \, (z \bar z)^{-i \, u} \left(\frac{z}{\bar z}\right)^{\frac{a}{2} - k} \, (\psi^1)^{a-k} (\psi^2)^k
\eeq
because $L$ (the Cartan generator of the Lorentz transformation) attributes weight $1,-1$ to $\psi^1, \psi^2$, respectively. Our immediate aim is to re-sum the infinite series in $z,\bar z$ that the weight factor creates; for now we turn a blind eye on all transformations required to rotate the hexagons in the internal space. This is why $J$ may be ignored for now, as well as additional weights $W(\alpha,\bar \alpha)$ accounting for to the $R$-charge of the states. We can also momentarily send $\psi^{a-k-1} \, \rightarrow \, \psi^{a-k}$ etc. since these are constant shifts. On the other hand, the summation ranges for the $k$-counter in (\ref{sl2BoundBits}) must be respected to obtain sensible results.

On the left and the right hexagon there is only one bound state and thus no scattering. Yet, the contraction rule for the outer hexagons enforces the scattering on the middle tile to be diagonal. Further, let us choose $3 \gamma, \, 1 \gamma$ kinematics on the middle hexagon in which case the scalar factor $h$ from \cite{Basso:2015zoa} becomes
\beq
h(u^{3 \gamma}, v^\gamma) \, = \, \Sigma(u^\gamma, v^\gamma)
\eeq
with the ``improved'' BES dressing phase \cite{Beisert:2006ez,Arutyunov:2009kf} in the mirror-mirror kinematics
\beq
\Sigma^{ab} \, = \,  \frac{\Gamma[1 + \frac{a}{2} + i \, u]}{\Gamma[1 + \frac{a}{2} - i \, u]} \, \frac{\Gamma[1 + \frac{b}{2} - i \, v]}{\Gamma[1 + \frac{b}{2} + i \, v]} \, \frac{\Gamma[1 + \frac{a+b}{2} - i (u-v)]}{\Gamma[1 + \frac{a+b}{2} + i (u-v)]} \, + O(g) \, . \label{sig11}
\eeq

A comprehensive discussion of the bound-state $S$-matrix is given in \cite{Arutyunov:2009mi}, although in the ``symmetric representation'' in which the r\^ole of bosons and fermions is exchanged. Hence in (\ref{sl2BoundBits}) one would literally swap $\phi \, \leftrightarrow \, \psi$). By way of example, we consider the scattering of two states of the first type in (\ref{sl2BoundBits}), both with $\alpha=1$ or both with $\alpha=2$. The relevant scattering matrix is called $\cX^{kl}_n(a,u,b,v)$ in \cite{Arutyunov:2009mi}. Here we associate $a,k,u,x^\pm(u)$ with the first state and $b,l,v,y^\pm(v)$ with the second. At bound-state length one (fundamental particles) one has $k \, = \, l \, = \, 0$ so that the complicated part of the $S$-matrix (see \eqref{defX} below) reduces to one. This describes the scattering of two equal fundamental fermions and
in agreement with the nomenclature of \cite{Beisert:2005tm} the remaining overall factor is called $D$.
The entire $S$-matrix can be changed by an overall factor, and indeed this $D$ is equal to the $A$-element in \cite{Beisert:2005tm}\footnote{It follows from here that the bound state length 1 part of the $S$-matrix of \cite{Arutyunov:2009mi} is in fact the inverse of that derived in \cite{Beisert:2005tm}.}.

We repeated the steps of \cite{Arutyunov:2009mi} to re-derive the $S$-matrix in the antisymmetric representation. Flipping the statistics means exchanging Poincar\'e and conformal supersymmetry, and also Lorentz and internal symmetry generators. Sticking to the same algebra conventions one obtains a sign flip on the rapidity parameters, so in particular $x^\pm \, \leftrightarrow \, x^\mp$. The $\cX$-element at bound state length one now describes the scattering of two equal bosons. We observe that what was called $D$ before now becomes $A^{-1}$. Hence for the antisymmetric representation the construction yields $S^{-1}$ without any rescaling.

Next, by observation --- at least in $3\gamma, \, 1\gamma$ kinematics and at leading order in $g$ --- the diagonal elements of our $S^{-1}$ in the antisymmetric representation are related to those of \cite{Arutyunov:2009mi} by flipping the sign of the rapidities, which has the interpretation of a complex conjugation or of taking a second inverse. In fact, this statement is true up to some global factors $(-1)^F$ which will be absorbed in the contraction prescription on the hexagons. Hence we can use the $S$-matrix of \cite{Arutyunov:2009mi} for our purposes, without any changes. 

Below we reproduce the expression for $\cX$ from \cite{Arutyunov:2009mi}, because it is the only one for which the original work gives a completely explicit writing. The other matrix elements are written as a sum over $x$'s with slightly shifted counters with certain matrices built from $x^\pm, y^\pm,a,b,k,l,n$ as coefficients, so they are similar albeit more complicated. We present some explicit expressions in Appendix \ref{app:A}.
\begin{equation}
 \label{defX} 
\begin{aligned}
\cX_n^{k,l} = &\, D \ \frac{\prod_{j=1}^n (a-j) \, \prod_{j=1}^{k+l-n} (b-j)}{\prod_{j=1}^k (a-j) \, \prod_{j=1}^l (b-j) \, \prod_{j=1}^{k+l} (-i \, \delta + \frac{a+b}{2} - j)}\\
& *\sum_{m=0}^k \left(\begin{array}{c} k \\ k-m \end{array} \right) \left(\begin{array}{c} l \\ n-m \end{array} \right) \, \prod_{j=1}^m c^+_j \, \prod_{j=1-m}^{l-n} c^-_j \, \prod_{j=1}^{k-m} d_{k-j+2} \, \prod_{j=1}^{n-m} \tilde d_{k+l-m-j+2}  \, ,
\end{aligned}
\end{equation}
where $\delta \, = \, u - v$ and
\beq
c^\pm_j = -i \, \delta \, \pm \frac{a-b}{2} - j + 1 \, , \quad d_j =\frac{a + 1 - j}{2} \, , \quad \tilde d_j = \frac{b + 1 - j}{2} \, .
\eeq
Further,
\beq
D \, = \, \frac{x^- -y^+}{x^+ - y^-} \sqrt{\frac{x^+}{x^-}} \sqrt{\frac{y^-}{y^+}} \quad
\stackrel{3 \gamma, \, 1 \gamma}{\longrightarrow} \quad
 - \frac{u^- -v^+}{u^+ - v^-} \frac{\sqrt{u^+ u^-} \sqrt{v^+ v^-}}{u^- v^+} \, + \, O(g^2) \, . \label{D31}
\eeq
At fixed bound state lengths $a,b$, let us denote the tensor product of two states of the first type listed in (\ref{sl2BoundBits}) as $|k,l\ra$. Their scattering takes the form
\beq
\mathbb{S} \, |k,l\ra \, = \, \sum_{n=0}^{k+l} \, \cX^{k,l}_n \, |n,k+l-n\ra \label{opS11}
\eeq
where $n \, = \, k$ is diagonal, so we adopt the convention in which the state with bound state length $a$ is written on the left before and after scattering. Suppose we label the states in \eqref{sl2BoundBits} as
\beq
| 1 \ra, \, | 2 \ra , \, |-\ra, \, |12\ra
\eeq
according to which scalar constituents they have. We will also usually drop the label $\cdot^{k,l}_k$ which may be readily re-instated.\footnote{We keep it only in Appendix \ref{app:D} to be closer to the notation of \cite{Arutyunov:2009iq}.}  Symbolically, equation \eqref{opS11} becomes $\mS \, |1,1 \ra \, = \, \cX \, |1,1\ra$. The complete set of diagonal matrix elements in our computation is
\begin{equation}
\begin{aligned}
&\mS \, |1,- \ra  \!&=&\;  \cY_{11} \, |1,-\ra \, , &&\qquad \mS \, |-,1\ra \!&=&\; \, \cY_{22} \, |-,1\ra \, ,\\
&\mS \, |1,12 \ra \!&=&\; \cY_{33} \, |1,12\ra \, , &&\qquad \mS \, |12,1 \ra\! &=&\; \cY_{44} \, |12,1\ra \, ,\\
&\mS \, |-,- \ra  \!&=&\;  \cZ_{11} \, |-,-\ra \, , &&\qquad \mS \, |-,12 \ra  \!&=&\;  \cZ_{22} \, |-,12\ra \, ,\\
&\mS \, |12,- \ra  \!&=&\; \, \cZ_{33} \, |12,-\ra \, , &&\qquad \mS \, |12,12 \ra \!&=&\; \, \cZ_{44} \, |12,12\ra,\\
&\mS \, |1,2 \ra  \!&=&\;  \cZ_{55} \, |1,2\ra \, , &&\qquad \mS \, |2,1 \ra \!&=&\; \, \cZ_{66} \, |2,1\ra \, . 
\end{aligned}
\label{defY}
\end{equation}
There is a second copy of $\tilde \cX$ algebraically equal to $\cX$ for the scattering of the states $|2,2\ra$ and in the same way $\tilde \cY$ (identical to $\cY$) is defined as in \eqref{defY} but for single scalars of flavour~two.

Any virtual particle used in the gluing procedure of \cite{Basso:2015zoa} is endowed with a ``mirror measure'' which comprises a factor $(g^2)^{l + 1}$. Here $l$ is the ``width" of the edge crossed by the virtual particle, i.e. the number of propagators forming that edge. At $O(g^2)$ the order estimate of \cite{Eden:2018vug} only allows gluing over edges of width zero. For the pentagon frame in Figure \ref{fig:2} these are the edges 13 or 14 or both. In the latter case one obtains $g^4$ from the measure factors, while the bound state $S$-matrix on the central hexagon contains terms of order $1/g^2$.

By way of example, including the two measure and the two weight factors as well as the dressing phase, the scattering process involving the $\cX$-matrix yields the sum-integral
\beq
I(\cX) \, = \, \sum_{a,b = 1}^\infty \sum_{k,l = 0}^{a-1,b-1} \, \int \frac{du \, dv\, a \, b \, g^4}{4 \pi^2 (u^2 + \frac{a^2}{4})^2 (v^2 + \frac{b^2}{4})^2} \, W_1 \, W_2 \, \Sigma^{ab} \, \cX_k^{k,l} \, .\nonumber
\eeq
This is a priori an $O(g^4)$ contribution, so it should drop from the one-loop result. However, we expect the scattering of the scalar constituents of the bound states to introduce braiding factors like
\beq
e^{i \frac{p_1}{2}} \, e^{-i \frac{p_2}{2}} \ \stackrel{3\gamma, \, 1\gamma}{\rightarrow} \ {\frac{\sqrt{u^+ u^-} \sqrt{v^+ v^-}}{g^2}} \label{xMom}
\eeq
where we have scaled back to the field theory convention of (\ref{defXRoot}) to meet the weak-coupling expansion. Importantly, this factor does not only adjust the leading power in the coupling constant to $g^2$, but it also removes the square-root branch cuts that would render inefficient the residue theorem as a means of evaluating the integrals over the rapidities $u,v$. In \cite{Fleury:2017eph} an ``averaging prescription'' for such additional braiding factors was suggested. With the present article we hope to shed some light on the question whether this prescription is the only possible one. 

Despite its appearance, the $\cX$-matrix has singularities in $\delta$ only in the lower half-plane. Poles in $\cX$ can therefore be avoided simply by closing the integration contour over the upper half-plane for $u$ and the lower half-plane for $v$. Doing so, the poles $u^-, \, v^+$ from the measure can contribute. Likewise, in the numerator of the phase, $\Gamma[1 + \frac{a}{2} + i \, u]$ and $\Gamma[1 + \frac{b}{2} - i \, v]$ develop singularities. Note however that we cannot localise both rapidities by poles from the phase:
\beq
u \, = \, i \, \left(m + \frac{a}{2}\right), \ v \, = \, - i \, \left(n + \frac{b}{2} \right) \quad \Rightarrow \quad \Gamma[1 + \frac{a+b}{2} + i (u-v) ] \, = \, \Gamma[1 - m - n]
\eeq
for $m,n \, \in \, \mathbb{N}$ so that this denominator $\Gamma$-function creates a zero in these cases. Thus at least one pole, perhaps a higher one, must come from the measure. Then, e.g. with $u \, = \, i \frac{a}{2}$,
\beq
\Sigma^{ab} \, = \, \frac{\Gamma[1] \, \Gamma[1 + \frac{b}{2} - i \, v] \, \Gamma[1 + a + \frac{b}{2} + i \, v]}{\Gamma[1+a] \, \Gamma[1 + \frac{b}{2} + i \, v]  \, \Gamma[1 + \frac{b}{2} - i \, v]}
\eeq
and therefore the term in the phase that could create a pole at $v_n =  - i (\frac{b}{2}+n)$ actually drops. One might be tempted to conclude that only the poles in the measure contribute to the sum of residua; in \cite{deLeeuw:2019tdd} we presented work in progress on this assumption. However, a derivative from a double pole at $u \, = \, i \frac{a}{2}$ acting upon the incriminated denominator term creates a factor $\psi(1 + \frac{b}{2} - i \, v)$ so that the phase can now have poles at $v_n, \, n \in \mathbb{N}$. There are similar residua with the roles of $u,v$ exchanged. In conclusion, we get to analyse the three cases
\beq
2 \, i (-u,v) \, \in \, \{ (a, b), \, (a, b + 2 \, n), \, (a + 2 \, n, b) \} \, . \label{threeRes}
\eeq
This list is complete due to the numerator $u^- - v^+$ of $D$ which suppresses one double pole of the factor $1/(u^- v^+)^2$ arising from the measure and the momentum dressing (\ref{xMom}). In Table~\ref{tab:diagonal} and Table~\ref{tab:2} we show the product of all $u^\pm, v^\pm$ factors for the various matrix elements. There is always only one double pole, whereby the above reasoning about the contributing residua is universal.

We should mention that the $\cZ$-elements have simple poles at $\delta  =  0$. Choosing the integration contour in the upper half-plane for $u$ and in the lower for $v$ this could just about contribute when $u \, = \, v \, = \, 0$, which is possible in, say, a principle-value prescription. Yet, this provides one pole while two are required to localise the two integrations, whereas no other term is singular at this point. 

\section{Re-summation of the residua at the poles of the measure}
\label{sec:2}

At leading order in the coupling in $3 \gamma, 1 \gamma$ kinematics, the diagonal $S$-matrix elements factor out a product of $u^\pm, v^\pm$. Importantly for our residue strategy, also a $(u^+ u^- - v^+ v^-)$ denominator in the $\cZ$-elements cancels which would otherwise lead to an expansion in terms of fractional powers of the cross-ratios. Writing the rapidities in terms of $\delta \, = \,  u - v, \, s \, = \, u + v$, we can completely factor out the dependence on $s$.

Further, localising both rapidities at the poles of the measure, so $u \, = \, i \frac{a}{2}, \, v \, = \, - i \frac{b}{2} \, \Rightarrow \, \delta = \frac{i}{2} (a+b)$, the remaining $\delta$-dependent part of most of the diagonal matrix elements reduces to a simple ratio of $\Gamma$-functions of the counters. In all these cases one finds the same kernel
\beq
K \, = \, \bino{k+l}{l} \bino{(a-k)+(b-l)}{a-k} \left[ \bino{a+b}{a}\right]^{-1} \, , \label{kernel}
\eeq
though with simple index shifts. We can express these functions as a product $f(a,k,b,l) \, K$. Conveniently, the dressing phase \eqref{sig11}
\beq
\Sigma^{ab} |_{u \, = \, i \frac{a}{2}, \, v \, = \, - i \frac{b}{2}} \, = \, \bino{a+b}{a} + \ldots
\eeq
 compensates the denominator of (\ref{kernel}). 

In all the matrix elements behaving in this way, the numerator $(u^- - v^+)$ of the universal pre-factor $D$ in (\ref{D31}) is compensated by a corresponding pole in the complicated part of $S$-matrix. We therefore obtain a non-vanishing result, in which the denominator factor $u^+ - v^- \rar a + b$ has been absorbed into $K$.

In Table~\ref{tab:diagonal} we list the factor $f$, the dressing by $P_i \, = \, e^{\frac{i}{2} p_i}$ necessary to obtain a rational integrand of order $O(g^2)$, and the total $s$-dependent part arising from this dressing, the measure, and the matrix elements including $D$.
\begin{table}[t]
\centering
\begin{tabular}{c | c | c | c}
$\cY,\cZ$ & $f$ & $P$ & $\prod u^\pm v^\pm$  \\
 \hline
$\cY_{11}$ & $- i \frac{a-k}{a (a-k+b-l)}$ & $1/P_2$ & $\frac{1}{(u^-)^2 u^+ v^- v^+}$ \\
$\cY_{22}$ & $i \frac{b-l}{b (a-k+b-l)}$ & $P_1$ & $\frac{1}{u^- u^+ v^- (v^+)^2}$ \\
$\cZ_{11}$ & $-\frac{(a l-b k)^2}{a^2 b^2 (k+l) (a-k+b-l)}$ & 1 & $\frac{1}{u^- u^+ v^- v^+}$ \\
$\cZ_{22}$ & $-\frac{l (b-l)}{b^2 (k+l) (a-k+b-l)}$ & 1 & $\frac{1}{u^- u^+ (v^+)^2}$ \\
$\cZ_{33}$ & $-\frac{k (a-k)}{a^2 (k+l)(a-k+b-l)}$ & 1 & $\frac{1}{(u^-)^2 v^- v^+}$ \\
$\cZ_{55}^-$ & $\frac{l (a-k)}{a b (k+l) (a-k+b-l)}$ & $1/(P_1 P_2)$ & $\frac{1}{(u^-)^2 u^+ v^-}$ \\
$\cZ_{55}^+$ & $\frac{l (a-k)}{a b (k+l) (a-k+b-l)}$ & $P_1 P_2$ & $\frac{1}{u^- v^- (u^+)^2}$ \\
$\cZ_{66}^-$ & $\frac{k (b-l)}{a b (k+l) (a-k+b-l)}$ & $1/(P_1 P_2)$ & $\frac{1}{(u^-)^2 u^+ v^-}$ \\
$\cZ_{66}^+$ & $\frac{k (b-l)}{a b (k+l) (a-k+b-l)}$ & $P_1 P_2$ & $\frac{1}{u^- v^- (v^+)^2}$
\end{tabular}
\caption{The diagonal elements of the bound-state $S$-matrix that factor at the poles of the measure, and their dressing by $u^\pm,v^\pm$.}
\label{tab:diagonal}
\end{table}

Our choice of $P_1 \, = \, \sqrt{u^+ u^-}/g, \, P_2 \, = \, g/\sqrt{v^+ v^-}$ decorations is motivated by removing or doubling square root factors in such a way that total order $O(g^2)$ is obtained. We can in fact understand the list in Table~\ref{tab:diagonal} as associating a $P_i$ factor with every bound state with a single scalar, see Section \ref{sec:2}. The fact that bound states with two scalar constituents $\phi^1 \phi^2$ do not create extra $P_i$ dressing is also apparent from our comments on the ``opposite wrapping'' calculation in \cite{Basso:2015zoa} by virtual magnons, cf.\ Appendix~\ref{app:D} and the conclusions.

There are two choices marked as $\pm$ for $\cZ_{55},\cZ_{66}$. These are related by a factor $P_1^2 P_2^2 \, = \, (u^+ u^-)/(v^+ v^-) + O(g^2)$. Further multiplication with this combination would neither change the order in $g$ nor re-introduce cuts. Yet, the listed possibilities are singled out by yielding maximally one double pole in $u^-$ or $v^+$ (not both), whereas all other choices yield higher singularities. When using the residue theorem on a higher pole one partially integrates derivatives onto the rest of the integrand. Each derivative creates a ``large logarithm'' since $\partial_u (z_1 \bar z_1)^{- i u} \, = \, - i \log(z_1 \bar z_1) (z_1 \bar z_1)^{- i u}$ and similarly for $v$. In off-shell kinematics (no Mandelstam invariant is identically zero due to light-like separation of some points) we should find one large logarithm per loop order, so that at $O(g^2)$ in fact only single derivatives are permitted. In test calculations with poles $(u^-)^n$ or $(v^+)^n$, $n < -2$ we obtained terms of transcendentality weight greater than two in the part without large logarithms, too, so when the derivatives fall upon $1/u^+,1/v^-$ or the scattering matrix. In conclusion, the $P_i$ decorations above should constitute an exhaustive list.

When the derivative acts on $(z_1 \bar z_1)^{-i u}$ or $(z_2 \bar z_2)^{-i v}$, the large logarithm it creates is a constant factor w.r.t. the quadruple sum over $a,k,b,l$, which must have polylogarithm weight one. In the cases in Table~\ref{tab:diagonal}, and also for the diagonal $\cX$-elements, this sum can be computed in closed form as we shall explain below. Our techniques also apply to the analytic evaluation of those terms in which the derivative acts on $1/u^+$ or $1/v^-$. Linear combinations of weight two hyperlogarithms are found. 

On the other hand, when the derivative acts on the $\delta$-dependent part of the matrix elements it destroys the factorisation properties yielding the kernel $K$. Yet, we could fit all of the resulting series on an ansatz built from the functions found in those terms amenable to explicit integration.

Further comments concern the matrix elements that do not factor at all: $\cX, \, \cY_{33}, \, \cY_{44},$ $ \cZ_{44}$. For the $\cX$-matrix we achieved an integration in closed form in all three situations, so with the derivative acting on the $(z \bar z)^{\ldots}$ factors, the other poles in the measure, and the matrix itself. This is possible because \cite{Arutyunov:2009mi} gives a completely explicit writing of $\cX$. The relevant computations are given in Appendix \ref{app:B}. For $\cY_{33},\cY_{44},\cZ_{44}$ we used a fit at polylogarithm level one, which could in turn be integrated in closed form for the terms with differentiated $1/u^+,1/v^-$. The series from derivatives of the $\delta$-part of the matrix elements could be fitted on the aforementioned ansatz. In future work we will use the simplifications of the $S$-matrix given in Appendix \ref{app:A} to derive complete results for the diagonal $\cY$ elements, and eventually try to address $\cZ$, too.

In all of $\cX, \, \cY_{33}, \, \cY_{44}, \, \cZ_{44}$ the numerator factor $u^- - v^+$ from $D$ is not cancelled. We put it into the collection of $u^\pm,v^\pm$ factors, while we attribute the non-singular denominator $u^+ - v^-$ to the $\delta$-part.
\begin{table}[t]
\centering
\begin{tabular}{c | c | c}
$\cX,\cY$ & $P$ & $\prod u^\pm v^\pm$  \\
 \hline
$\cX$ & $P_1/P_2$ & $\frac{u^- - v^+}{(u^-)^2 u^+ v^- (v^+)^2}$ \\
$\cY_{33}$ & $1/P_2$ & $\frac{u^- - v^+}{(u^-)^2 u^+ (v^+)^2}$ \\
$\cY_{44}$ & $P_1$ & $\frac{u^- - v^+}{(u^-)^2 v^- (v^+)^2}$ \\
$\cZ_{44}$ & $1$ & $\frac{u^- - v^+}{(u^-)^2 (v^+)^2}$ \\
\end{tabular}
\caption{Reduced table for the non-factoring diagonal matrix elements.}
\label{tab:2}
\end{table}
One of the two derivatives from the residua has to act on the numerator $u^- - v^+$ to obtain a non-vanishing contribution. Alternatively, we may expand the $u^\pm, v^\pm$ factors into two terms, each displaying one double pole.

Last, we remark that there is no distinction between the two copies of the $\cX,\cY$ matrices, respectively, w.r.t. the comments in this section.

\subsection{Integrating $\cY_{11}$} 

The matrix $\cY_{11}$ occurs in the scattering of $\phi^1 (\psi^1)^{a-1-k} (\psi^2)^k$ over
$(\psi^1)^{b-l} (\psi^2)^b$. We put aside the factor $\sqrt{\bar z_1/z_1}$ that arises from the action of the $L$ generator as well as a similar factor $\sqrt{\alpha_1/\bar \alpha_1}$ from the action of the $R$ on the scalar in the first bound state. How to attribute $R$-charge to a given scattering process is 
discussed in Section \ref{sec:2}. Finally, we use the notation $\{z_1, \bar z_1, z_2, \bar z_2\} \rar \{z_1, b_1, 1/a_2, 1/y_2\}$ for ease of reading.

In $3 \gamma, 1\gamma$ kinematics the $\cY_{11}$ element is of leading order $1/g$, further $1/P_2$ is also $O(g^{-1})$ and the measure contributes a factor $g^4$. Next, we pick residua at $u^-$ (double pole) and $v^-$, whereby $1/(u^+ v^-) \rar 1/(a b)$, which cancels against the numerator of the measure. Integrating the derivative onto $(z_1 b_1)^{-i u}$ one obtains a global factor $-i \log(z_1 b_1)$. From $f \, k \, \Sigma$ as listed in Table~\ref{tab:diagonal} we find the quadruple sum
\begin{equation}
- \frac{S_{\log}(Y_{11})}{\log(z_1 b_1)} \, = \, \sum_{a,b,k,l} z_1^{a-k} \, b_1^k \, y_2^{b-l} \, a_2^l  \, \frac{\quad \Gamma[a-k+b-l] \qquad  \Gamma[1+k+l]}{a \, \Gamma[a-k] \, \Gamma[1 + b - l] \ \Gamma[1+k] \,  \, \Gamma[1+l]} \label{ylSer}
\end{equation}
where $a,b \, = \, 1 \ldots \infty, \, k,l \, = 0 \ldots a-1,b$. To obtain independent sums we would like to shift $a \rar a+k, \, b \rar b+l$, but the explicit factor $1/a$ is a hinderance. We can eliminate it by differentiation in the absolute value of $z_1$: Defining
\begin{equation}
r^2 \, = \, z_1 b_1 \, , \quad p^2 \, = \, \frac{z_1}{b_1} \qquad \Rightarrow \qquad r \, \frac{\partial}{\partial r} \, z_1^{a-k} b_1^k \, = \, a \; z_1^{a-k} b_1^k \, . \
\end{equation}
The inverse operation is $\int dr/r$, which is, of course, only defined up to a function of the phase $z_1/b_1$ and the other two variables $y_2,a_2$. Differentiating and shifting the sums we obtain
\begin{equation}
- r \frac{\partial}{\partial r} \frac{S_{\log}(\cY_{11})}{\log(z_1 b_1)} \, = \, \sum_{a=1}^\infty \sum_{k=0}^\infty \left( \sum_{b=1}^\infty \biggl|_{l=0}+ \sum_{b=0}^\infty \sum_{l=1}^\infty\right)
z_1^a y_2^b \bino{a+b-1}{a-1} \, b_1^k a_2^l \bino{k+l}{k} \label{splitSum}
\end{equation}
The sums are of geometric type and can easily be taken in closed form:
\beq
- r \frac{\partial}{\partial r} \frac{S_{\log}(\cY_{11})}{\log(z_1 b_1)} \, = \, \frac{z_1 \, (a_2 + y_2 - a_2 y_2 - b_1 y_2 - a_2 z_1)}{(1 - b_1) (1 - a_2 - b_1) (1 - z_1) (1 - y_2 - z_1)} \label{yrSer}
\eeq
Expressing the r.h.s. in terms of modulus and phase of $z_1$ we can apply the integral operator $\int dr/r$ which, by definition, adds a letter to a hyperlogarithm. The result is
\begin{equation}
\label{oneResult} 
\begin{aligned}
- \frac{S_{\log}(\cY_{11})}{\log(z_1 b_1)}  = & \frac{z_1 \, (\log[1 - b_1] - \log[1 - z_1])}{(b_1 - z_1)} \\
&+  \frac{z_1 \, (\log[1 - a_2] - \log[1 - a_2 - b_1] - \log[1 - y_2] + \log[1 - y_2 - z_1])}{(b_1 - z_1 - b_1 \, y_2 + a_2 \, z_1)} 
\\
&+  c(z_1/b_1,y_2,a_2) 
\end{aligned}
\end{equation}
where $c$ is a fairly complicated expression constant in $r$. Re-expanding the result of the integration and comparing to the original series we see that $c \rar 0$ is the right result, so the entire constant part has to be subtracted. This happened in all instances in which we applied the differentiation/integration trick, presumably for the reason that all the denominators in the set of functions do depend on the modulus $r$ in question. It would be important to better understand this point.

We can thus eliminate $a$ from the denominator of (\ref{ylSer}) without loss of information. The root of the procedure is then a rational function (\ref{yrSer}) and we add polylogarithm levels by the integration in the modulus $r$. For instance, the contribution from the derivative falling onto $1/u^+$ is as the r.h.s. of (\ref{ylSer}) but with a second factor of $1/(i a)$. To find it, we can apply the operation $\int dr/r$ a second time, on (\ref{oneResult}):
\beq
S_\mathrm{mes}(\cY_{11}) \, = \, - \frac{z_1 \, (\mathrm{Li}_2[b_1] - \mathrm{Li}_2[z_1])}{(b_1 - z_1)} + \frac{z_1 \, \left(\mathrm{Li}_2\left[\frac{b_1}{1-a_2}\right] - \mathrm{Li}_2\left[\frac{z_1}{1-y_2}\right]\right)}{(b_1 - z_1 - b_1 \, y_2 + a_2 \, z_1)}
\eeq
At this point we cannot present an algorithm for the evaluation of the remaining contribution, say, $S_\mathrm{mat}(\cY_{11})$ in which the scattering matrix (including the phase) is differentiated. The situation is strictly analogous for all other entries in Table~\ref{tab:diagonal} barring for $\cZ_{11}$ which is special in that there is no double pole. The integration algorithm described here can therefore catch the entire contribution from $\cZ_{11}$. We dedicate the next section to this computation. 

As in \cite{Fleury:2017eph} we assumed that factors like $|u|, |v|$ do not arise from the $x(u)$ functions or the expansion of $\Sigma^{ab}$, and that $(-1)^a, (-1)^b, (-1)^{a b}$ are unphysical and have to be undone by the contraction prescription on the hexagons.

\subsection{Integrating $Z_{11}$}
\label{subsec:2:2}

Picking residua from $1/(u^- v^+)$ we have $a b / (u^+ v^-) \rar 1$ as before. We find the three sums
\begin{equation}
\begin{aligned}
\{ S_1, \, S_2, \, S_3 \} = & \sum_{a,b \, = \, 1}^\infty \sum_{k,l \, = \, 0}^{a,b} z_1^{a-k} \, b_1^k \, y_2^{b-l} \, a_2^l \, \frac{\Gamma[k+l] \Gamma[a-k+b-l]}{\Gamma[1+a-k] \Gamma[1+b-l]} \\ & * \left\{ - \frac{k}{a^2 \, \Gamma[k] \Gamma[1+l]}, \, \frac{2}{a \, b \, \Gamma[k] \Gamma[l]}, \, - \frac{l}{b^2 \, \Gamma[1+k] \Gamma[l]} \right\}  \label{threeSums}
\end{aligned}
\end{equation}
To apply the trick of the last section we will remove $\{a^2,\, a \, b, \, b^2\}$, respectively, from the denominators by differentiation intending to integrate twice in the modulus of $z_1$ or $y_2$, or mixed in the middle term. 

After differentiation we first have to execute the sums, though. In the process one notices that the cases $b \, = \, l$ and $a \, = \, k$ are special and have to be dealt with separately. (Simultaneously, these conditions would make the summand vanish.) The two special cases yield
\begin{equation}
\begin{aligned}
S_4  = & - \sum_{a,b \, = \, 1}^\infty \sum_{k \, = \, 0}^\infty  z_1^{a - k} \, b_1^k \, a_2^b \frac{(a - k) \Gamma[b + k]}{a^2 \, \Gamma[1 + b] \Gamma[1 + k]} \, , \\
S_5  =&  - \sum_{a,b \, = \, 1}^\infty  \sum_{l=0}^\infty b_1^a \, y_2^{b-l} \, a_2^l \frac{(b-l) \Gamma[a + l]}{b^2 \,  \Gamma[1 + a] \Gamma[1 + l]} \, ,
\end{aligned}
\end{equation}
where we remove the explicit factors $1/a^2, 1/b^2$ by differentiation. Henceforth we may assume $k \, < \, a$ and $l \, < \, b$ which also simplifies shifting the sums. With these definitions
\begin{eqnarray}
S_1 & = & -\frac{(1-a_2) \, b_1}{(1-a_2-b_1)^2} (\log[1-y_2] + \log[1-z_1] - \log[1-y_2-z_1]) \, , \nonumber \\
S_4 & = & \frac{z_1}{(1-z_1)^2} (\log[1-a_2-b_1] - \log[1-b_1]) \, , \nonumber \\
S_2 & = & \frac{2 \, a_2 \, b_1}{(1-a_2-b_1)^2} (\log[1-y_2] + \log[1-z_1] - \log[1-y_2-z_1]) \, , 
\label{S15Z}
\end{eqnarray}
\begin{eqnarray}
S_3 & = & -\frac{a_2 \, (1-b_1)}{(1-a_2 - b_1)^2} (\log[1-y_2] + \log[1-z_1] - \log[1-y_2-z_1]) \, ,\nonumber \\
S_5 & = & \frac{y_2}{(1-y_2)^2} (\log[1-a_2-b_1] - \log[1-a_2]) \, . \nonumber
\end{eqnarray}
Interestingly, we do not find rational expressions with polynomial denominators to the power of one, but rather sums of logarithms divided by the square of a polynomial. In fact, the first integration in the modulus (which can be $|z_1|$ for $S_1,S_4,S_2$ and $|y_2|$ for $S_3,S_5,S_2$) yields single logarithms over polynomials to the first power, as in the problem discussed in the last section. However, although the structure is the same, the logarithms now come with various rational numerator factors so that these expressions are more lengthy than above. We refrain from giving them here and rather state the end result, i.e. the sum of all doubly integrated $S_i$:
\begin{equation}
\label{sZ11}
\begin{aligned}
S(\cZ_{11}) = & - \frac{y_2 \, \mathrm{Li}_2\left[ a_2\right] - a_2 \, \mathrm{Li}_2\left[ y_2 \right]}{a_2 - y_2}  - \frac{z_1 \, \mathrm{Li}_2\left[ b_1\right] - b_1 \, \mathrm{Li}_2\left[ z_1 \right]}{b_1 - z_1} - \frac{(b_1 y_2 + a_2 z_1) \, L_{\cZ_{11}}}{b_1 y_2 - a_2 z_1} \\ & + \frac{(1 - b_1) y_2 \, \mathrm{Li}_2\left[ \frac{a_2}{1 - b_1}\right] - a_2 (1 - z_1) \,\mathrm{Li}_2\left[ \frac{y_2}{1 - z_1}\right]}{a_2 - y_2 + b_1 y_2 - a_2 z_1} \\
& + \frac{(1 - a_2) z_1 \, \mathrm{Li}_2\left[ \frac{b_1}{1 - a_2}\right] - b_1 (1 - y_2) \, \mathrm{Li}_2\left[ \frac{z_1}{1 - y_2}\right]}{b_1 - z_1 -b_1 y_2 + a_2 z_1}
\end{aligned}
\end{equation}
In the last formula, the pure function\footnote{A linear combination of hyperlogarithms with constant coefficients} $L_{\cZ_{11}}$ is:
\begin{equation}
\label{zComp}
\begin{aligned}
L_{\cZ_{11}} & =
\log[1 - a_2]^2 - \log[1 - b_1]^2 - \log[1 - a_2] \log[1 - y_2] + \log[1 - b_1] \log[1 - z_1] \\[1.5 mm]
&+ \log[1 - a_2 - b_1] (-\log[1 - a_2] + \log[1 - b_1] + \log[1 - y_2] - \log[1 - z_1]) \\[1.5 mm]
&+\mathrm{Li}_2\left[ a_2 \right]-\mathrm{Li}_2\left[ y_2 \right] -\mathrm{Li}_2\left[ b_1\right]+\mathrm{Li}_2\left[ z_1\right] \\
& -\mathrm{Li}_2\left[ \frac{a_2}{1 - b_1}\right] +\mathrm{Li}_2\left[ \frac{y_2}{1 - z_1}\right]
+ \mathrm{Li}_2\left[ \frac{b_1}{1 - a_2}\right]-\mathrm{Li}_2\left[ \frac{z_1}{1 - y_2}\right] \\ &+\mathrm{Li}_2\left[ -\frac{a_2 - y_2}{1 - a_2} \right] - \mathrm{Li}_2\left[ -\frac{b_1 - z_1}{1 - b_1}\right] \\
& - \mathrm{Li}_2\left[ -\frac{a_2 - y_2 + b_1 y_2 - a_2 z_1}{(1 - a_2 - b_1) (1 - z_1)}\right]
+\mathrm{Li}_2\left[ -\frac{b_1 - z_1 - b_1 y_2 + a_2 z_1}{(1 - a_2 - b_1) (1 - y_2)}\right]
\end{aligned}
\end{equation}

\vskip 0.15 cm

\subsection{The non-factoring matrix elements and educated guesses}

For $S(\cX)$ we can obtain all three pieces $S_\mathrm{log}, S_\mathrm{mes}, S_\mathrm{mat}$ in closed form, see Appendix \ref{app:B}. In particular,
\beq
\frac{S_\mathrm{log}(\cX)}{\log[z_1 b_1 \, y_2 a_2]} \, = \, \frac{y_2 z_1 (\log\left[\frac{(1 - a_2)(1 - b_1)}{1 - a_2 - b_1}\right] - \log\left[\frac{(1 - y_2)(1 - z_1)}{1 - y_2 - z_1}\right]}{a_2 b_1 - a_2 b_1 y_2 - a_2 b_1 z_1 - y_2 z_1 + a_2 y_2 z_1 + b_1 y_2 z_1} \label{sLogX}
\eeq
and $S_\mathrm{mes}(\cX)$ contains two pure functions (whose difference is closely related to $L_{\cZ_{11}}$) over the denominators
\beq
\{ a_2 b_1 - a_2 b_1 y_2 - y_2 z_1 + a_2 y_2 z_1, a_2 b_1 - a_2 b_1 z_1 - y_2 z_1 + b_1 y_2 z_1\} \, ,
\eeq
respectively. The total set of denominators (disregarding the intermediate stage of the $\cZ_{11}$ computation given in equation (\ref{zComp})) is
\begin{eqnarray}
&& \{ a - y, \ b - z, \ b \ y - a \, z, \, a - y + b \, y - a \, z, \, b - b \, y - z + a \, z, \label{dens} \\
&& \phantom{\{} a \, b - a \, b \, y - a \, b \, z - y \, z + a \, y \, z + b \, y \, z,  \, a \, b - a \, b \, y - y \, z + a \, y \, z, \ a \, b - a \, b \, z - y \, z + b \, y \, z \} \nonumber
\end{eqnarray}
where we have dropped the $1,2$ subscripts, because the sums have all been done whence there can be no confusion between $a_2, b_1$ and the bound state lengths $a,b$ which have disappeared from the expressions. So far, the logarithmic functions seen at transcendentality weight one are
\beq
\{\log[1-a], \, \log[1-b], \, \log[1-a-b], \, \log[1-y], \, \log[1-z], \, \log[1-y-z] \} \, . \label{weight1}
\eeq
Importantly, all the functions that we have found by integration have the following features:
\begin{itemize}
\item They are sums $\sum_i f_i/d_i$ with $d_i$ in the set (\ref{dens}).
\item Every numerator $f_i$ is a sum of rational factors times pure functions, where the rational factors are terms of the denominator $d_i$.
\item At weight one the pure functions are linear combinations of the logarithms in (\ref{weight1}).
\end{itemize}
We can now write an ansatz using the most general linear combination of denominator terms times logarithms in the numerator of every $d_i$. This is constrained by two requirements:
\begin{itemize}
\item If $l_i$ is the linear combination acting as a numerator for $d_i$, the Taylor expansion\footnote{A well-defined Taylor series is obtained introducing an order parameter $\{a,b,y,z\} \rar \{ o \, a, o \, b, o \, y, o \, z\}$ and expanding in $o$.} of $l_i/d_i$ must not have any polynomial denominator.
\item Allowing all terms of each denominator $d_i$ as rational factors in the numerator $l_i$ obviously allows for a complete cancellation of $d_i$, which makes the eight parts linearly dependent. To avoid this we do not allow the respective last denominator term in any numerator but $l_1$.
\end{itemize}
Indeed, the transcendentality one part of $S(\cY_{33}),S(\cY_{44}),S(\cZ_{44})$ --- on which we cannot run our integration scheme as even the undifferentiated matrix elements do not factor into $\Gamma$-functions --- fits such an ansatz! We find
\begin{eqnarray}
- \frac{S_\mathrm{log}(\cY_{33})}{\log[a \, b \, y \, z]} & = & \frac{a \, y \, z \left(\log\left[ \frac{(1-a)(1-b)}{1-a-b} \right] - \log\left[ \frac{(1-y)(1-z)}{1-y-z}\right]\right)}{a \, b - a \, b \, y - a \, b \, z - y \, z + a \, y \, z + b \, y \, z} \, , \nonumber \\
S_\mathrm{log}(\cY_{44}) & = & \frac{b}{a} \, S_\mathrm{log}(\cY_{33}) \, , \\
\frac{S_\mathrm{log}(\cZ_{44})}{\log[a \, b \, y \, z]} & = & \frac{(1-a-b) \, y \, z \log\left[ \frac{(1-a)(1-b)}{1-a-b} \right] + (1-y-z) \, a \, b  \log\left[ \frac{(1-y)(1-z)}{1-y-z}\right]}{a \, b - a \, b \, y - a \, b \, z - y \, z + a \, y \, z + b \, y \, z} \, . \nonumber
\end{eqnarray}
To compute $S_\mathrm{mes}(\cY_{33})$ and $S_\mathrm{mes}(\cY_{44})$ from these formulae one straightforwardly proceeds by integration; for $\cZ_{44}$ there is no such contribution due to the simple structure of the $u^\pm, v^\pm$ factor in Table~\ref{tab:2}. On the other hand, $S_\mathrm{mat}(\ldots)$ remains inaccessible once again.

Nevertheless, the explicit weight two results found so far --- in particular $S(\cZ_{11})$ as stated in (\ref{zComp}), $S_\mathrm{mes}(\cX)$ and $S_\mathrm{mat}(X)$ ---- offer a range of hyperlogarithms that we can express by a set of basis functions with the help of the symbol \cite{Goncharov:1998kja,Goncharov:2010jf}. The letters in their symbols are
\beq
\{ 1 - a, \ 1 - b, \ 1 - a - b, \ 1 - y, \ 1 - z, \ 1 - y - z \} \label{firstEntry}
\eeq
and
\beq	
\{ a, \ b, \ y, \ z, \ a - y, \ b - z, \
a - y + b \, y - a \, z, \ b - b \, y - z + a \, z,  \ a \, b - a \, b \, y - a \, b \, z - y \, z + a \, y \, z + b \, y \, z \} \,  .
\eeq
Since we obtained the functions from a Taylor expansion for small $\{ a, b, y, z \}$ all symbols obey a first entry condition: only the letters in the set (\ref{firstEntry}) occur in the first slot of the symbols. A basis of functions is given by
\begin{itemize}
\item products of any two of the simple logarithms in (\ref{weight1}), yielding 21 functions,
\item the eight dilogarithms in the third and fourth line of (\ref{zComp}),
\item the four dilogarithms in the last two rows of (\ref{zComp}), supplemented by the fifth variant
\beq
\mathrm{Li}_2\left[ \frac{a \, b - a \, b \, y - a \, b \, z - y \, z + a \, y \, z + 
  b y z}{(1 - a) (1 - b) (1 - y - z)} \right] \, .
\eeq
\end{itemize}
We can separately fit every $S_\mathrm{mat}(\ldots)$ contribution onto an ansatz of the same type as for the weight one problem, though employing the 34 weight two functions just listed.

\subsection{The forgotten residue comes to our aid}

A cornerstone of the hexagon tessellation approach is ``embedding invariance'' \cite{Eden:2016xvg,Fleury:2016ykk}, i.e. tiling independence of the results. For every pentagon frame we have the right to choose any tiling; in particular, for the pentagon skeleton with vertices 12345 we may choose to triangulate as in Figure \ref{fig:4} in Section \ref{sec:tessel5}. Clearly, it is also allowed to turn the corner --- for us $x_1$ --- at which the six hexagons touch to any of the other vertices. Hence the result of our computation must be invariant under cyclic shifts of the pentagon frame. Yet, cyclic symmetry (\ref{cycZ1}) relates the last three functions of our basis to the first two with inverted $\{a,b,y,z\}$, so
\beq
\mathrm{Li}_2\left[ - \frac{a-y}{(1-a) \, y} \right] \, , \qquad \mathrm{Li}_2\left[ - \frac{b-z}{(1-b) \, z} \right] \, . 
\eeq
whose symbols contain the terms $- y \otimes y, - z \otimes z$, respectively, amongst others violating the first entry condition. In conclusion, all such functions must cancel from the sum over residua for the 12345 matter pentagon on its own, and indeed separately for any point-flipped version, too.

In a similar vein, all the denominators (\ref{dens}) arise from point permutations of $z-b$, but the cyclic shifts (\ref{cycZ1}) alone fail to generate the three cases
\beq
 \{ b \ y - a \, z, \, a \, b - a \, b \, y - y \, z + a \, y \, z, \ a \, b - a \, b \, z - y \, z + b \, y \, z \} \, .
\eeq

Everything falls into place upon taking into account the second type of residue, in which a derivative acts on the term $\Gamma[1 + (a + b)/2 + i (u - v)]$ in the denominator of the BES phase in mirror/mirror kinematics (\ref{sig11}). The $\psi$-function so created yields an extra pole with residue
\begin{eqnarray}
\mathrm{res}_{v \, = \, -i \left(\frac{b}{2} +n \right)} \; \partial_u \Sigma^{ab} |_{u \, = \, +i \frac{a}{2}} & = & \phantom{-} i \, \bino{a+b+n}{a}  \, , \\
\mathrm{res}_{u\, = \, +i\left(\frac{a}{2} + n\right)} \; \partial_v \Sigma^{ab} |_{v \, = \, -i\frac{b}{2}} & = & -i \, \bino{a+b+n}{b} \, , \nonumber
\end{eqnarray}
where $n \in \mathbb{N}$. The related Taylor series are similar to what was found before, but there is a fifth sum $\sum_{n=1}^\infty (a_2 y_2)^n \ldots$ or  $\sum_{n=1}^\infty (z_1 b_1)^n \ldots$, respectively. 

At these residua, none of the matrix elements factors, so that we can only proceed by fitting. The sums $S_{\psi}(\cX), \, S_\psi(\cY_{11})+S_\psi(\cY_{33}), \, S_\psi(\cY_{22})+S_\psi(\cY_{44}), \,  S_\psi(\cZ_{22}) + S_\psi(\cZ_{33}) + S_\psi(\cZ_{44}), \, S_\psi(\cZ_{55}^\pm), \, S_\psi(\cZ_{66}^\pm)$ can be matched by the ansatz used before. Adding in the previously discussed contributions $S_\mathrm{log}, \, S_\mathrm{mes}, \, S_\mathrm{mat}$ we obtain the results (\ref{firstRes}) \ldots (\ref{lastRes}) in Section \ref{sec:tessel5} in which all unwanted functions and denominators disappear. Adding the $\cY$ and $\cZ$ matrix elements as indicated is compatible with their $R$-charge and $e^{\frac{i}{2} p}$ dressing; in the final result they appear in the stated sums at any rate. A little unfortunately for our philosophy, the ansatz cannot match each $S_\psi(\cY)$ or $S_\psi(\cZ)$ sum on its own, so e.g. $S_\psi(\cY_{11})$ without adding $S_\psi(\cY_{33})$. It would be interesting to understand the relevant extension.

However, in constructing the ansatz we did not import knowledge from field theory: the function space was found from the solvable part of the problem. Below we will fix the $R$-charge assignments and thereby the final result of the calculation only from the requirement of triangulation invariance, here in the guise of cyclic shifts of the pentagon matter skeleton.

A central element in recent work on scattering amplitudes of the ${\cal N}=4$ theory is the knowledge of the relevant space of functions. While four- and five-point amplitudes are exactly given by the BDS ansatz \cite{Bern:2005iz}, starting from six points there is a ``remainder function'' depending on conformal cross ratios \cite{Drummond:2007bm}. In \cite{Golden:2013xva} it was suggested that the arguments of hyperlogarithms in the remainder functions can be inferred employing the ``transmutation moves'' of cluster algebras. Combining this idea with geometrical considerations, the function space for six- and seven-particle amplitudes has been understood well enough to permit fits up to rather higher order \cite{Dixon:2011pw,Dixon:2013eka,Dixon:2014voa,Caron-Huot:2016owq,Golden:2014xqa,Dixon:2016nkn}. We want to point out that the integration algorithm defined above could generate a structure like a cluster algebra: from one or several ``roots'' we obtain various hyperlogarithms and/or denominators in a way reminiscent of transmutation moves.

In the process of integrating $S(\cZ_{11})$ we have met an instance in which the root of the process is not rational: the expressions in (\ref{S15Z}) where of the type simple logarithm divided by the square of a polynomial. The logarithm series is the simplest instance of a hypergeometric sum that we may encounter in this context. In Appendix \ref{app:C} we present a control calculation on $S(\cZ_{11})$ which does not use the differentiation/integration trick outlined above. Its agreement with \eqref{sZ11},\eqref{zComp} confirms that there is no problem with integration constants in the original approach. In this second calculation we rather replace the hypergeometric series (two $_2F_1$ or one $_3F_2$) by their standard integral representations in terms of Euler kernels. Leaving the integrations for later, the integrands are again products of some (possibly composite) argument with the bound state counters as their exponents and $\Gamma$-functions as coefficients. Thus the procedure closes on hypergeometric sums. At the last stage, the integrations add polylogarithm weight.

In Appendix \ref{app:B} we use this second method more extensively to derive $S_\mathrm{log}(\cX)$ and $S_\mathrm{mat}(\cX)$. Once again, all contributions are term-wise of transcendentality two. 

\section{Charges and results}
\label{sec:3}

\subsection{Review of the one-magnon case at four points} \label{secOneMag}

In \cite{Fleury:2016ykk} gluing over the width zero diagonals of an empty square frame of propagators is discussed. Let us first discuss this on the tiling in the left panel of Figure 1. 

\begin{figure}[t]
\begin{center}
\includegraphics[width=\linewidth]{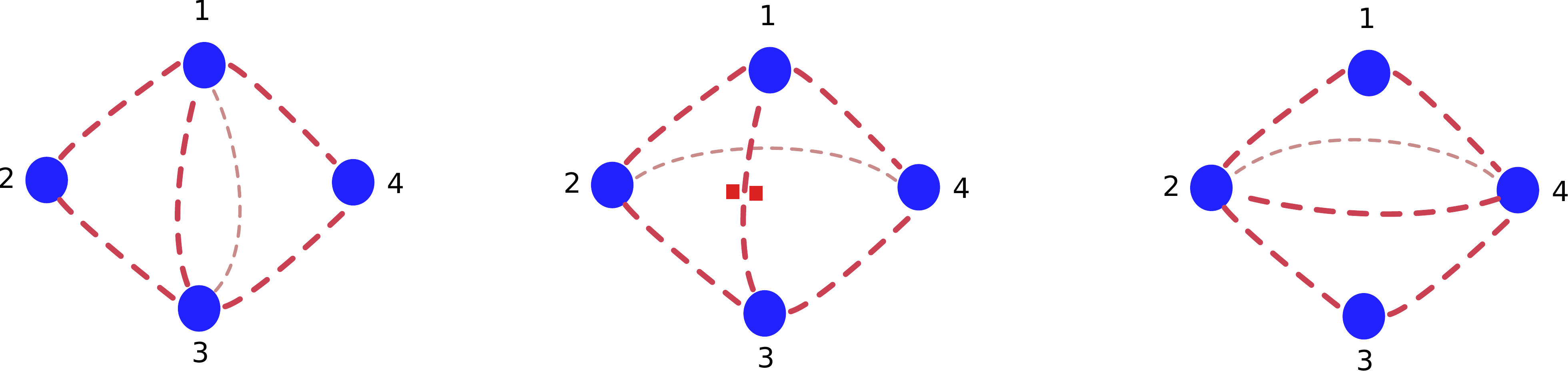}
\end{center}
\caption{Tilings for planar four-point functions: the left and the right panel show the two possible triangulations of the sphere (up to point permutations of the operators). In the middle we zoom in on gluing the two front hexagons by a single virtual magnon.}
\label{fig:3}
\end{figure}

In the middle panel we focus on gluing over the front diagonal. The rotation to general kinematics associates the variables
\beq
X \,= \, \frac{x_{14}^2 \, x_{23}^2}{x_{12}^2 \, x_{34}^2} \, =: \, z \, \bar z \, , \qquad Y \,= \, \frac{y_{14}^2 \, y_{23}^2}{y_{12}^2 \, y_{34}^2} \, =: \, \alpha \, \bar \alpha \,
\eeq
with the weight factor (\ref{rotateHex}). The one-magnon gluing over the diagonal on the back of the figure can be brought to the front by rotating through $\pi$ vertically or horizontally, which implies the point exchanges $1 \, \leftrightarrow \, 3$ or $2 \, \leftrightarrow \, 4$, respectively. 
Both operation send the cross ratios $X, \, Y$ on the r.h.s. of the last equation to their reciprocals, so the corresponding variable transformation is $z \, \leftrightarrow \, 1/z, \, \bar z \, \leftrightarrow \, 1/\bar z$ and $\alpha \, \leftrightarrow \, 1/\alpha, \, \bar \alpha \, \leftrightarrow \, 1/\bar \alpha$. On the tetrahedron type tiling in the right panel of Figure 1 gluing on the back of the figure is related to that on the front by the simultaneous variable exchange $x_1 \leftrightarrow x_2, \, x_3 \leftrightarrow x_4$ which acts on the cross ratios as the identity.

The variables $y_i$ in the definition of $Y$ are contracted onto the $R$-symmetry indices of the  ${\cal N}= 4$ scalars forming the four BPS operators. Without the projectors, BPS correlators must carry finite dimensional representations of the internal symmetry group. Upon projection this means that the entire correlation functions has to be a polynomial in $y_{ij}^2$, see \cite{Eden:2000qp} and references therein. In the case of minimal charge --- so length two --- at every point, the square ``skeleton diagramme''  or ``matter frame'' of propagators $y_{ij}^2/x_{ij}^2$ that we want to dress by virtual exchanges yields the numerator polyomial $y_{12}^2 \, y_{23}^2 \, y_{34}^2 \, y_{41}^2$. This can suppress the singularities in both $Y, \, 1/Y$, but not of higher powers thereof. We conclude that allowed terms in the gluing process are
\beq
f_1(z,\bar z) + f_2(z,\bar z) \, Y + f_3(z,\bar z) \frac{1}{Y} \, .
\eeq
In particular, we should only see single poles in $\alpha, \bar \alpha$.

For the bound state $\psi_1^{a-k} \, \psi_2^k$ without bosons we obtain the sum-integral 
\beq
I_1 \, = \, \sum_{a=1}^\infty \sum_{k=0}^a \int \frac{du}{2 \, \pi} (z \bar z)^{- i u} \left(\frac{z}{\bar z}\right)^{\frac{a}{2} - k} \frac{g^2 \, a}{(u + i \frac{a}{2})^2 (u - i \frac{a}{2})^2}
\eeq
For $|z| \, \leq \, 1$ we close the integration contour for $u$ over the upper half-plane thus picking the residue at the double pole at $u \, = \, i \, a/2$. This yields a Taylor-expansion in positive powers of $z, \bar z$ valid for $|z| \leq 1$:
\beq
I_1 \, = \,  - g^2 \, \sum_{a=1}^\infty \sum_{k=0}^a \, z^{a-k} \bar z^k \left[ \frac{\log(z \bar z)}{a} - \frac{2}{a^2} \right] \, = \, \frac{- g^2}{z-\bar z} \sum_{a=1}^{\infty} \left(z^{a+1} - \bar z^{a+1}\right) \left[ \frac{\log(z \bar z)}{a} - \frac{2}{a^2} \right] 
\eeq
and finally
\beq
I_1 \, = \, \frac{g^2}{z-\bar z} \left( z \, [2 \, \mathrm{Li}_2(z) + \log(z \bar z) \, \log(1 - z) ] - \bar z \, [2 \, \mathrm{Li}_2(\bar z) + \log(z \bar z) \, \log(1 - \bar z) ] \right) \, . \label{I1}
\eeq
For the bound state  $\phi_1 \phi_2 \, \psi_1^{a-1-k} \, \psi_2^{k-1}$ we find in very much the same way
\beq
I_2 \, = \,  \frac{g^2}{z-\bar z} \left( \bar z \, [2 \, \mathrm{Li}_2(z) + \log(z \bar z) \, \log(1 - z) ] - z \, [2 \, \mathrm{Li}_2(\bar z) + \log(z \bar z) \, \log(1 - \bar z) ] \right) \, , \label{I2}
\eeq
though here $0 \, < \, k \, < \, a$. Finally,
\beq
I_1 + I_2 \, = \, g^2 (z + \bar z) \, B(z,\bar z)
\eeq
with the Bloch-Wigner dilogarithm
\beq
B(z,\bar z) \, = \, \frac{1}{z - \bar z} \left[ 2 \, \mathrm{Li}_2(z) - 2 \, \mathrm{Li}_2(\bar z)+ \log(z \bar z) \, \log\left(\frac{1 - z}{1 - \bar z}\right) \right] \, . \label{bwdilog}
\eeq
Separately, $I_1, \, I_2$ are not proportional to this function due to the rational factors in the numerator. We find the individual terms of the denominator --- i.e. $z$ and  $\bar z$ --- multiplying distinct ``pure functions'' composed of generalised logarithms.

For the remaining cases $\phi_{1,2} \, \psi_1^{a-1-k} \, \psi_2^k$ we obtain a prefactor
\beq
\sqrt{\frac{\bar z}{z}} \sqrt{\frac{\alpha}{\bar \alpha}} \, , \quad \sqrt{\frac{\bar z}{z}} \sqrt{\frac{\bar \alpha}{\alpha}} \, ,
\eeq
respectively, from the offset 1 in the exponent of $\psi_1$ and the $R$-charge of the scalar fields. Summing $0 \, \leq \, k \, < \, a$ yields
\beq
I_3 \, = \, g^2 \sqrt{\frac{\alpha}{\bar \alpha}} \, \sqrt{z \bar z} \; B(z,\bar z) \, , \qquad I_4 \, = \, I_3 \, : \, \alpha \, \leftrightarrow \, \bar \alpha \, .
\eeq
Note that the sum-integral itself produces a rational factor $z/(z-\bar z)$ in front of the pure functions, so once again one of the terms of the denominator occurs also in the numerator. Note further that the $\alpha, \bar \alpha$ dependence splits the entire computation into parts that individually yield $B(z,\bar z)$, so a single-valued combination of hyperlogarithms.

According to \cite{Fleury:2016ykk}, $I_3, \, I_4$ should receive an additional minus sign, and we should act on the bound states with the charge $J$, whose eigenvalues on the scalar constituents $\phi^a$ can be
\begin{equation}
\sqrt{\frac{\alpha \bar \alpha}{z \bar z}} \, , \qquad \sqrt{\frac{z \bar z}{\alpha \bar \alpha}} \, .
\end{equation}
Expecting $J$ to measure spin-chain length, its action on $\phi^1$ and $\phi^2$ should be equal. However, it must in fact be opposite so that there is no net effect on $\phi^1 \phi^2$ in the same bound state, as we can also see in the transfer matrix exercise in Appendix C. In the present context, the sum of $I_1$ and $I_2$ cannot yield a single-valued function if there is an $\alpha$-dependent relative factor\footnote{In the vein of what is said below, adding the gluing processes over the back and the front of the figure does not help, because the two ``halves'' of the Bloch-Wigner function in $I_1, \, I_2$ do not separately have a simple transformation law under  inversion of $z, \bar z$.}. For an opposite action there are the two possibilities
\begin{eqnarray}
\oplus & : & J \, \phi^1 \, = \, Z \, \phi^1 \, , \quad J \, \phi^2 \, = \, Z^{-1} \, \phi^2 \, , \qquad Z \, = \, \sqrt{\frac{\alpha \bar \alpha}{z \bar z}} \, ,  \label{defJ} \\
\ominus & : & J \, \phi^2 \, = \, Z \, \phi^2 \, , \quad J \, \phi^1 \, = \, Z^{-1} \, \phi^1 \, .
\end{eqnarray}
Collecting terms and omitting the overall factor of $g^2$ we obtain
\begin{eqnarray}
\oplus & : & n^+ \, = \, \left[ z + \bar z - \alpha - \frac{z \bar z}{\alpha} \right] \, B(z, \bar z) \, , \\
\ominus & : & n^- \, = \, \left[ z + \bar z - \bar \alpha - \frac{z \bar z}{\bar \alpha} \right] \, B(z, \bar z) \, .
\end{eqnarray}
Both of these go into themselves under the inversion of $z, \alpha$. Since $\alpha, \bar \alpha$ are root-functions of the $y_{ij}^2$ it seems hard to avoid the conclusion that only the weighted sum
\beq
M^{(1)} \, = \, \frac{1}{2} [n^+ + n^- ] \, = \, \frac{1}{2} [\, m(z,\alpha) + m(1/z,1/\alpha)\,] \, , \qquad m(z,\alpha) \, = \, (z + \bar z - \alpha - \bar \alpha) \, B(z,\bar z) \, . \label{ms}
\end{equation}
can be the physical answer, in which the two combinations
\beq
\alpha + \bar \alpha \, = \, \frac{y_{12}^2 y_{34}^2 - y_{13}^2 y_{24}^2 - y_{14}^2 y_{23}^2}{y_{12}^2 y_{34}^2} \, , \qquad \frac{1}{\alpha} + \frac{1}{\bar \alpha} \, = \, \frac{y_{12}^2 y_{34}^2 - y_{13}^2 y_{24}^2 - y_{14}^2 y_{23}^2}{y_{14}^2 y_{23}^2}
\eeq
are both allowed by harmonic analyticity. Further, $z \leftrightarrow \bar z$ and $\alpha \leftrightarrow \bar \alpha$ symmetry must appear by reparametrisation invariance. It would be nicer if this could be made manifest by a tiling argument. Last,
averaging over a charge assignment seems an unusual procedure, and indeed in the two-magnon-computation described below it is not necessary.

\subsection{Five points at $O(g^2)$: three tiles glued by two virtual particles}
\label{sec:tessel5}

We consider the correlator of five length two half-BPS operators. The tree level defines two types of skeleton diagrams:
\beq
(12)^2 \, (34)(45)(53)\, , \qquad (12)(23)(34)(45)(51)
\eeq
(and their permutations) where a propagator $y_{ij}^2/x_{ij}^2$ between points $i,j$ is denoted as $(ij)$. In a  field theory computation the disconnected first ``matter frame'' would not receive a one-loop correction: virtual contributions within the two- and three-point blocks, respectively, cancel due to non-renormalisation theorems for BPS operators \cite{DHoker:1998vkc,Eden:1999gh}, and any virtual exchange between the two frames has vanishing colour factor at least at one loop. In \cite{Fleury:2016ykk} this was pronounced as a selection rule against ``one-edge-reducible'' graphs; in \cite{Eden:2017ozn} we gave a proof based on the properties of one-loop colour factors.

\begin{figure}[t]
\begin{center}
    \includegraphics[width=0.3\linewidth]{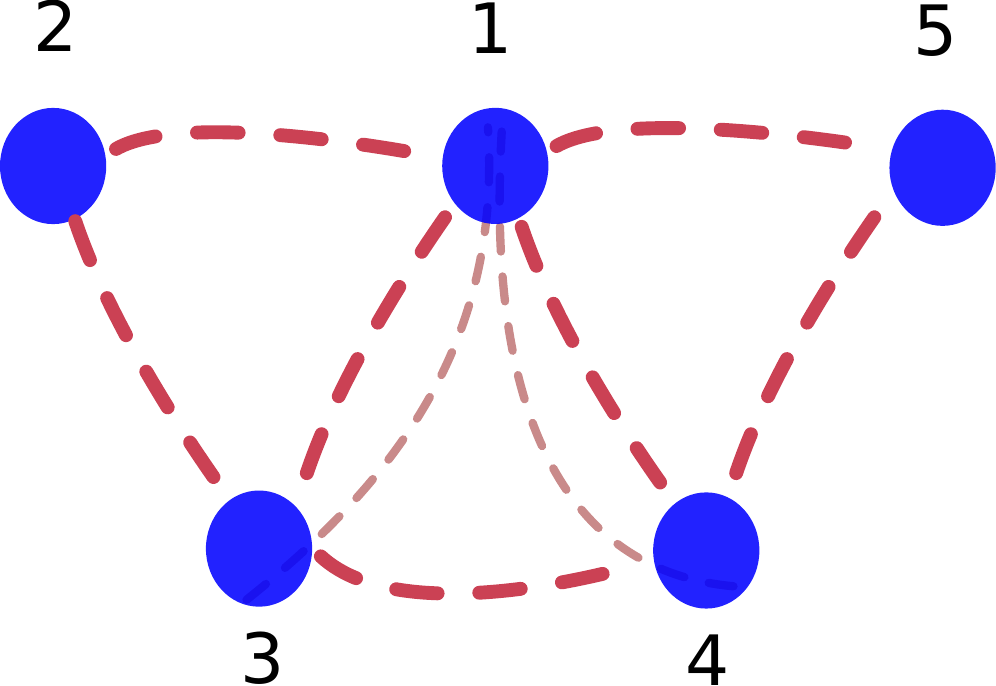}
\end{center}
\caption{One of four possible tessellations of the pentagon skeleton diagram by six hexagons}
\label{fig:4}
\end{figure}

On the other hand, in field theory as well as in the integrability calculation using tilings, the pentagon frame must be dressed by virtual particles. In the latter case, the result must be independent of the tiling. We will choose the tessellation in Figure \ref{fig:4}, leaving a more complete analysis for future work. Having selected the type of tessellation (there are three other possibilities related by flipping diagonals on squares), triangulation indepedence \cite{Eden:2016xvg,Fleury:2016ykk} still imposes the invariance of the result under cyclic rotations of $12345$. On either side of the figure we have two one-magnon exchanges of the type discussed in the last section:

\begin{figure}[t]
\begin{center}
\includegraphics[width = \linewidth]{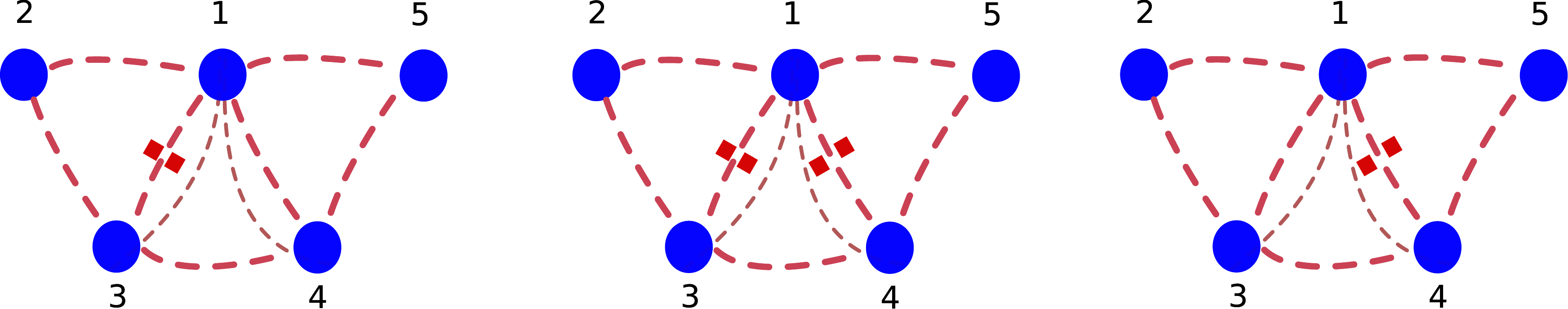}
\end{center}
\caption{The three $O(g^2)$ gluing processes on the front of Figure \ref{fig:4}}
\label{fig:5}
\end{figure}

The two magnon-process in the middle panel of Figure \ref{fig:5} is the focus of interest in \cite{Fleury:2017eph}. In the present article we not only ask the question of how to re-sum the corresponding residue calculation, but putting together the results for the various scattering processes we want to study possible implications for the formalism, which is the object of this section.

Our first comments will regard the variables in this exercise. A full-blown conformal five-point function will depend one five cross ratios. To be able to use the same four Cartan generators for the rotation of ``canonical hexagons'' \cite{Fleury:2016ykk} to four-point kinematics for either one-magnon exchange we restrict to kinematics on a complex line (we display the spatial 1,2 plane, the other coordinates are zero) $x_1 \, = \, \{0,0\}, \, x_3 \, = \, \{0,\infty\}, \, x_4 \, = \, \{0,1\}$, and
\beq
\begin{aligned}
x_2 & = \, \left\{ \frac{i}{2}\left(\frac{1}{z_1}-\frac{1}{\bar z_1}\right), \frac{1}{2}\left(\frac{1}{z_1}+\frac{1}{\bar z_1}\right) \right\} \, , \\ \qquad  x_5 & = \, \left\{ \frac{i}{2} \left(\frac{z_2}{z_2-1} - \frac{\bar z_2}{\bar z_2 - 1} \right), \frac{1}{2} \left( \frac{z_2}{z_2-1} + \frac{\bar z_2}{\bar z_2 - 1} \right) \right\}
\end{aligned}
\eeq
with which
\beq
X_1 \, = \, \frac{x_{14}^2 x_{23}^2}{x_{12}^2 x_{34}^2} \, = \, z_1 \bar z_1 \, , \qquad X_2 \, = \, \frac{x_{15}^2 x_{34}^2}{x_{13}^2 x_{45}^2} \, = \, z_2 \bar z_2
\eeq
and similarly for internal cross ratios $Y_1 \, = \, \alpha_1 \bar \alpha_1, \, Y_2 \, = \, \alpha_2 \, \bar \alpha_2$. For the one-magnon exchanges the internal space structures
\beq
\alpha_1 + \bar \alpha_1 \, = \, \frac{y_{12}^2 y_{34}^2 - y_{13}^2 y_{24}^2 - y_{14}^2 y_{23}^2}{y_{12}^2 y_{34}^2} \, , \qquad \frac{1}{\alpha_2} + \frac{1}{\bar \alpha_2} \, = \, \frac{y_{13}^2 y_{45}^2 - y_{14}^2 y_{35}^2 - y_{15}^2 y_{34}^2}{y_{15}^2 y_{34}^2}
\eeq
are allowed by the pentagon frame's numerator $y_{12}^2 y_{23}^2 y_{34}^2 y_{45}^2 y_{51}^2$. This occurs in $m(z_1,\alpha_1)$ and $m(1/z_2,1/\alpha_2)$, respectively. The full result $M^{(1)}$ in (\ref{ms}) cannot appear in the sum of processes because of the functions $m(1/z_1,1/\alpha_1)$ and $m(z_2,\alpha_2)$ which would lead to poles in $y_{ij}^2$. These functions in the one-magnon processes have to cancel against parts of the two-magnon gluing.

The double-gluing unites three tiles, and thus represents a five-point object, for which the four Cartan generators previously used for the rotation of the hexagons could prove insufficient to even reach kinematics on the complex line. Fortunately, we can rule out this difficulty in the present case: for our choice of points, the middle hexagon is canonical (though, w.r.t. one side in a different order of the points than w.r.t. the other). Thus a rotation concerning point 2 acts only on the left hexagon and a rotation of point 5 only the right hexagon, so that the two operations are in fact independent. The weight factors arises by definition from the states inserted on the rotated hexagons, so here those on the outer hexagons. The gluing on the right hexagon is therefore exactly as in the one-magnon case in Section \ref{secOneMag}, although the points 1234 are mapped to 1345. On the other hand, the gluing of the left hexagon seems of opposite orientation to the definition in Section \ref{secOneMag}. To complete the argument we observe that rotating the middle panel of Figure 1 corresponds to the simultaneous point exchange $1 \leftrightarrow 3, \, 2 \leftrightarrow 4$ which is an identity map on the cross-ratios. 

Next, gluing on the back of Figure \ref{fig:4} is related to that on the front by a rotation around the symmetry axis through the point $x_1$. This amounts to the simultaneous point exchange $2 \leftrightarrow 5,  \, 3 \leftrightarrow 4$ which induces the map $X_1 \leftrightarrow 1/X_2$ etc. or equivalently $z_1 \leftrightarrow 1/z_2, \, \bar z_1 \leftrightarrow 1/\bar z_2$ on the cross ratios. Note that this nicely agrees with the fact that we identified the $Y_1,Y_2$-dependence of  $m(z_1,\alpha_1)$ and $m(1/z_2,1/\alpha_2)$ as allowed in the end result. The complete set of five cyclic rotations map $z_1$ to
\beq
\left\{z_1,- \frac{(1-z_2)}{z_1 z_2}, -(1-z_1) z_2, \frac{1}{z_2}, \frac{1 - z_2 + z_1 z_2}{z_1} \right\} \label{cycZ1} 
\eeq
Now, in our attempt in Section \ref{sec:2} to resum the multiple hypergeometric series we found e.g. in the case of the $\cZ_{11}$-element of the scattering matrix
\begin{eqnarray}
(z_1 \bar z_1)^{-i u} \left(\frac{z_1}{\bar z_1}\right)^{\frac{a}{2}-k} (z_2 \bar z_2)^{-i v} \left(\frac{z_2}{\bar z_2}\right)^{\frac{b}{2}-l} \stackrel{u = i \frac{a}{2}, \, v = -i\frac{b}{2}}{\longrightarrow} z_1^{a-k} \, \bar z_1^k \, z_2^{-l} \, \bar z_2^{-b+l}
\end{eqnarray}
where we had redefined $ \{ z_1, \bar z_1, z_2, \bar z_2 \} \rightarrow \{ z, b, 1/a, 1/y \}$ to obtain a series of the type
\begin{eqnarray}
\sum_{a=1}^\infty \sum_{k=0}^a \sum_{b=1}^\infty \sum_{l=0}^b \, f(a,k,b,l) \, z^{a-k} \, b^k \, y^{b-l} \, a^l
\end{eqnarray}
in terms of four small variables. Why does this contain functions of the type $\log(1-a-b), \, \log(1-y-z)$ when we have conjugated at the second point, so $a = 1/z_2, y = 1/\bar z_2$, while the cyclic rotations transform $\{z_1,z_2\}$ and $\{\bar z_1, \bar z_2\}$ within the two sets, but do not mix them? We have associated the weight factor at point one with the state on the left hexagon, say, $X$. According to our understanding of the gluing procedure --- see Appendix C --- we should insert the same state on the central hexagon.
The  $3 \gamma$ kinematics means transporting the bound state on the left edge of the middle hexagon in the middle panel of Figure 3 to the $0 \gamma$ physical edge shrunken to a circle segment around point 4. While we keep the weight factor fixed, the actual bound state scattered changes from $X \rar - \bar X$.
On the right of the figure the opposite is true: let the weight factor be associated with a state $Y$ on both sides of the edge 14.
It is not changed by the $1\gamma$ rotation to the physical edge.
When choosing the variables we have conjugated at the second point, though, and not at the first. The net effect is $z \, \leftrightarrow \, \bar z, \, \alpha \, \leftrightarrow \, \bar \alpha$ at both points, which is clearly an unimportant change of parametrisation. As a confirmation of what has been said, note that the variable transformation for flipping back and front of the figure
\beq
\left\{ a \leftrightarrow b, \, y \leftrightarrow z, \, \alpha_1 \leftrightarrow \frac{1}{\bar \alpha_2}, \, \alpha_2 \leftrightarrow \frac{1}{\bar \alpha_1} \right\}
\eeq
sends the $\cX, \, \cZ_{11}, \, \cZ_{44}$ scattering processes to themselves and exchanges $\cY_{11} \leftrightarrow \cY_{22}, \, \cY_{33} \leftrightarrow \cY_{44}, \, \cZ_{22} \leftrightarrow \cZ_{33}, \, \cZ_{55}^\pm \leftrightarrow \cZ_{55}^\mp, \, \cZ_{66}^\pm \leftrightarrow \cZ_{66}^\mp $, where $\pm$ refers to decorations with momentum factors, see Section \ref{sec:2}.

Before we state the results for the various scattering processes, let us point out a degeneracy created by the restricted kinematics. If we write the five permutations of $B(z,b)$ induced by (\ref{cycZ1}) as $B_i \, = \, L_i/D_i$, where $D_i$ are the rational denominators and $L_i$ the pure functions, then
\beq
\sum_{i=1}^5 L_i \, = \, 0 \, .
\eeq
When expressing the re-summed matrix elements in terms of the $B_i$ we can always add in this vanishing linear combination. We use this freedom to write $\cZ_{11} + \cZ_{22} + \cZ_{33} + \cZ_{44}$ in a manifestly $z \leftrightarrow \bar z$ symmetric form, but otherwise leave it aside for the time being. Denoting the re-summed scattering processes as $S(\ldots)$ we find
\begin{eqnarray}
&& S(\cZ_{11}) + S(\cZ_{22}) + S(\cZ_{33}) + S(\cZ_{44}) \, = \, \frac{1}{2} \biggl[ -(b+z) B_1 - (a + y) B_4 \label{firstRes} \\ && + \left(\frac{1 - a}{b} + \frac{1 - y}{z} \right) B_2 + \left( \frac{1 - b}{a} +\frac{1 - z}{y} \right) B_3  + \left( \frac{-1 + a + b}{a b} + \frac{-1 + y + z}{y z} \right) B_5 \biggr] \, , \nonumber
\end{eqnarray}
\beq
S(\cX) \, = \, - \sqrt{\frac{\alpha_1}{\bar \alpha_1}} \sqrt{\frac{\bar \alpha_2}{\alpha_2}} \, \frac{1}{\sqrt{a y \, b z}} \, B_5 \, ,
\eeq
\beq
S(\cY_{11}) + S(\cY_{33}) \, = \, \sqrt{\frac{\alpha_1}{\bar \alpha_1}} \biggl[ - \sqrt{b z} \, B_1 + \frac{1}{\sqrt{b z}} \, B_2 + \frac{1}{\sqrt{b z}} \, B_5 \biggr] \, ,
\eeq
\beq
S(\cY_{22}) + S(\cY_{44}) \, = \, \sqrt{\frac{\bar \alpha_2}{\alpha_2}} \biggl[  - \sqrt{a y} \, B_4 + \frac{1}{\sqrt{a y}} \, B_3 + \frac{1}{\sqrt{a y}} \, B_5 \biggr] \, ,
\eeq
The expressions for $S(\tilde \cX), S(\tilde \cY)$ --- so for the second copy of $\cX,\cY$ for states with scalar composites of opposite flavour --- are obtained from the last three formulae by conjugating all $\alpha, \bar \alpha$ variables. Finally, 
\beq
S(\cZ_{55}^-) \, = \, \sqrt{\frac{\alpha_1}{\bar \alpha_1}} \sqrt{\frac{\alpha_2}{\bar \alpha_2}} \, \frac{\sqrt{a y}}{\sqrt{b z}} \, B_2 \, , \qquad S(\cZ_{55}^+) \, = \, \sqrt{\frac{\alpha_1}{\bar \alpha_1}} \sqrt{\frac{\alpha_2}{\bar \alpha_2}} \, \frac{\sqrt{b z}}{\sqrt{a y}} \, B_3 \, ,
\eeq
\beq
S(\cZ_{66}^-) \, = \, \sqrt{\frac{\bar \alpha_1}{\alpha_1}} \sqrt{\frac{\bar \alpha_2}{\alpha_2}} \, \frac{\sqrt{a y}}{\sqrt{b z}} \, B_2 \, , \qquad S(\cZ_{66}^+) \, = \, \sqrt{\frac{\bar \alpha_1}{\alpha_1}} \sqrt{\frac{\bar \alpha_2}{\alpha_2}} \, \frac{\sqrt{b z}}{\sqrt{a y}} \, B_3 \, . \label{lastRes}
\eeq
From these equations it is already apparent that we will need to introduce a relative minus sign between sums $S(\ldots)$ with and without $\alpha$ decoration in order to fall upon the $m(z,\alpha)$ functions. Doing so and imposing cyclic invariance on the sum of the one- and two-magnon processes on the front and the back of the figure, respectively, yields a system of equations that does not completely fix the solution: define
\beq
j_1 \, = \, \frac{\sqrt{\alpha_1 \bar \alpha_1}}{\sqrt{b z}} \, , \qquad j_2 \, = \, \sqrt{a y} \sqrt{\alpha_2 \bar \alpha_2} \, .
\eeq
We find that all $S(\ldots)$ can occur in exactly one decoration with these eigenvalues of the $J$-charge:
\beq
c_X \, \frac{j_2}{j_1} \, S(\cX) \, , \quad c_{\tilde \cX} \, \frac{j_2}{j_1} \, S(\tilde X) \, , \label{res1}
\eeq
\beq
c_{Y13} \, \frac{1}{j_1} \, [ S(Y_{11}) + S(Y_{33}) ] \, , \qquad c_{\tilde Y24} \, j_2 \, [ S(\tilde Y_{22}) + S(\tilde Y_{44}) ] \, ,
\eeq
\beq
c_{\tilde Y13} \, \frac{1}{j_1} \, [ S(\tilde \cY_{11}) + S(\tilde \cY_{33}) ] \, , \qquad c_{Y24} \, j_2 \, [ S(\cY_{22}) + S(\cY_{44}) ] \, ,
\eeq
\beq
c_{Z5-} \, \frac{1}{j_1 \, j_2} \, S(\cZ_{55}^-) \, , \qquad c_{Z5+} \, j_1 \, j_2 \, S(\cZ_{55}^+) \, ,
\eeq
\beq
c_{Z6-} \, \frac{1}{j_1 \, j_2} \, S(\cZ_{66}^-) \, , \qquad c_{Z6+} \, j_1 \, j_2 \, S(\cZ_{66}^-) \, .
\eeq
The constants have to obey the equations
\beq
c_X + c_{\tilde X} \, = \, c_{Y13} + c_{\tilde Y24} \, = \, c_{\tilde Y13} + c_{Y24} \, = \, c_{Z5-} + c_{Z5+} \, = \,  c_{Z6-} + c_{Z6+} \, = \, 1 \, . \label{gradingEqs}
\eeq
These conditions connect pairs of constants for quantities related by flipping the back and the front of the figure. The averaging procedure in \cite{Fleury:2017eph} would put all constants to $1/2$. We see that there are many more possibilities; in particular in the pairs one can choose $(c,c') \, = \, (0,1)$, which opens the possibility to assign only one $e^{\frac{i}{2} p}$ decoration to $Z_{55}$ and $Z_{66}$. The $O(g^2)$ end result for dressing the tiling in Figure 2 with virtual magnons does not depend on the particular values of the constants as long as (\ref{gradingEqs}) is fulfilled. It is given by the cyclic sum
\beq
m(z_1,\alpha_1) + \mathrm{cyclic}
\eeq
where the transformation of the variables is as in (\ref{cycZ1}). Although this is perhaps difficult to infer from the explicit $\alpha$-dependence, the very fact that the first term generates all others by cyclicity ensures harmonic analyticity in all five parts.

In Table~\ref{tab:3} below we list the contributing $J$-charge assignments juxtaposed to the decoration of the $S$-matrix elements by the $P_j \, := \ e^{\frac{i}{2} p_j}$ factors needed to make them contribute at $O(g^2)$ and/or yield homogeneous polylogarithm level two. The indexing in the first two columns is similar to that in formula \eqref{defY}: - for no scalar or two in the bound state, 1 for $\phi^1$, 2 for $\phi^2$. In the last two columns $O(M), \, O(P)$ we display the leading perturbative order of the bound state matrix element and of the momentum dressing, respectively.
\begin{table}\centering
\begin{tabular}{ c | c | c | c | c | c | c | c }
$X$ & $Y$ & $S$ & $P$ & $J$ & $\oplus,\ominus$ & $O(M)$ & $O(P)$  \\
 \hline
 - & 1 & $\cY_{22}$ & $P_1$ & $j_2$ & $\oplus$ & $1/g$ & $1/g$ \\
 - & 2 & $\tilde \cY_{22}$ & $P_1$ & $j_2$ & $\ominus$ & $1/g$ & $1/g$ \\
1 & - & $\tilde \cY_{11}$ & $1/P_2$ & $1/j_1$ & $\ominus$ & $1/g$ & $1/g$ \\
2 & - & $\cY_{11}$ & $1/P_2$ & $1/j_1$ & $\oplus$ & $1/g$ & $1/g$ \\
1 & 1 & $\cZ_{66}^-$ & $1/(P_1 P_2)$ & $1/(j_1 j_2)$ & $\ominus$ & $1/g^2$ & $1$ \\
1 & 1 & $\cZ_{66}^+$ & $P_1 P_2$ & $j_1 j_2$ & $\oplus$ & $1/g^2$ & $1$ \\
1 & 2 & $\tilde \cX$ & $P_1/P_2$ & $j_2/j_1$ & $\ominus$ & $1$ & $1/g^2$ \\
2 & 1 & $\cX$ & $P_1/P_2$ & $j_2/j_1$ & $\oplus$ & $1$ & $1/g^2$ \\
2 & 2 & $\cZ_{55}^-$ & $1/(P_1 P_2)$ & $1/(j_1 j_2)$ & $\oplus$ & $1/g^2$ & $1$ \\
2 & 2 & $\cZ_{55}^+$ & $P_1 P_2$ & $j_1 j_2$ & $\ominus$ & $1/g^2$ & $1$
\end{tabular}
\caption{$J$ and $e^{\frac{i}{2}p}$ decorations in the $3 \gamma, 1 \gamma$ kinematics.}
\label{tab:3}
\end{table}
Here, $\cY_{11}$ is a short for $S(\cY_{11})+S(\cY_{33})$ and similarly for the other fixed combinations. The first two columns list the bound states $X,Y$ inserted.
Now, according to the crossing rules on the hexagon \cite{Basso:2015zoa}, the $3 \gamma$ rotation causes the conjugation $X \, \rightarrow \, -\bar X$ while $Y$ is not touched by the $1 \gamma$ rotation. This is why, for instance, the scattering process in the fifth row yields a $\cZ_{66}$ element instead of $\cX$. The overall order of the terms listed here is $1/g^2$ in all terms; another factor $g^4$ comes from the mirror measure for the two bound states.

Very clearly, in the contributing terms $P_1$ follows $j_2$ and $P_2$ follows $j_1$. This rule is \emph{almost} as found in \cite{Fleury:2017eph}, see the next subsection. We can raise a question mark about the interpretation, though: in all the pairs of constants in equation (\ref{gradingEqs}), one pertains to the $\oplus$ grading and the other to $\ominus$. Thus in our computation there is no need to average over gradings. From the current set of data we cannot reach a final conclusion; in this respect it will certainly be helpful to adapt this computation to the other tessellations of the pentagon frame, and eventually to the other $\gamma$ rotation.

A side effect of the averaging scenario is that it introduces higher order terms that may or may not be needed: By way of example, $S(\cX)$ with the $\ominus$ grading can contribute at $O(g^2)$. If the $\oplus$ grading is also used in the way defined in \cite{Fleury:2017eph} one obtains a term of total order (and presumably also polylogarithm level) six. 

\subsection{Comments on other mirror rotations}

Instead of shifting the magnons to the physical edge at point 4 by the mirror transformations
$X \, Y \rar - \bar X(3 \gamma) \, Y(1 \gamma)$ we can also use $X \, Y \rar Y(1 \gamma) \, X(-3 \gamma)$. In this kinematics the scalar $h$ factor is of order $1/g^2$, so combined with the measure the pre-factors are $O(g^2)$. The matrix-part of the scattering and the $e^{\frac{i}{2} p}$ dressing do not have a monodromy, and therefore their kinematics is effectively $1 \gamma$ rotated at both points. The order of the $S$-matrix components is
\beq
1 \gamma, \, 1 \gamma \ : \ O(\cX) \, = \, 1, \quad O(\cZ_{11}) \, = \, 1, \quad O(\cZ_{55}), O(\cZ_{66}) \, = \, g^2, \quad O(\cY_{ii}) \, = \, g \, . \label{order11}
\eeq
Hence in order to find an $O(g^2)$ contribution, in this kinematics $\cY$ all need to be dressed by $1/P$, while $\cX$ requires $P_1/P_2$ or $P_2/P_1$ and $\cZ_{55}, \cZ_{66}$ need $1/(P_1 P_2)$, so one has two choices for $\cX$ and one for $\cZ_{55}, \cZ_{66}$. Using the interpretation given above, i.e. to attribute the entire weight factor (including the $R$-charge) and also the $J$ weight on the edge where the states are inserted, we have to modify the momentum dressing as $P_1 \rar 1/P_1$ w.r.t.~Table~\ref{tab:3}. Interestingly, there are once again two possible momentum decorations $P_1/P_2, \, P_2/P_1$ when the bound states $X,Y$ both have one scalar constituent of flavour 1, or when both have one scalar of flavour 2. Finally, the choice $X \, Y \rar Y(5 \gamma) \, X(1 \gamma)$ of \cite{Fleury:2017eph} ought to be strictly equivalent by cyclicity of the central hexagon.

Sticking to the momentum dressing in Table~\ref{tab:3} we would assign $P_1/ P_2$, which is now of order $g^0$ to $\cZ_{55},\cZ_{66}$, so that these parts of the calculation are switched off at $O(g^2)$. Worse, $\cX$ could now pick up $(P_1 P_2)^{\pm 1}$ both of which are not of the right order; in particular, the choice with exponent -1 is an $O(g^0)$ term, and so a tree contribution from virtual magnons. All other momentum dressings have respected our worst case estimate \cite{Eden:2018vug} for the perturbative order at which dressing by virtual magnons comes in.

In fact, \cite{Fleury:2017eph} does not mention such difficulties. Translated into our conventions, their rule exactly amounts to evaluating $R$- and $J$-charge as we defined, but to assign the inverse momentum dressing w.r.t.\ $P_1$. A static link between $J$-charge and momentum dressing cannot exist. Is it therefore necessary to attribute momentum dressing once the magnons have been rotated to one physical edge? 

\section{Conclusions}

Our derivation of the result for the two-magnon gluing depicted in Figure \ref{fig:2} is built on inherent features of the hexagon tessellation approach. It matches that of the original article \cite{Fleury:2017eph}, and with it the field theory calculation\footnote{Note that the Lagrangian insertion technique for ${\cal N} = 2$ superfields \cite{Howe:1999hz} could have given the same result much earlier at little expense. However, the first explicit writing of one-loop five-, six- and seven-point half-BPS correlation functions computed by this means appeared in \cite{Alday:2010zy} (and hence later than in \cite{Drukker:2008pi}) when the correlator/Wilson-loop duality spurred interest in higher-point functions.} \cite{Drukker:2008pi}.

In \cite{Fleury:2017eph} the scattering on the central hexagon in Figure \ref{fig:2} is evaluated in different kinematics (so by moving the particles onto one physical edge in a different way). The identity
\beq
\la \mathfrak{h} | X(u^{6 \gamma}) Y(v) \ra \, = \, - \la \mathfrak{h} | Y(v) \bar X(u) \ra
\eeq
met by fundamental magnons can only hold for a bound state $X(u)$ when the integral over its rapidity $u$ and the sums over the two counters (the length and the number of spinor indices 
with value two) are taken. On the integrand/summand level our $\bar X(u^{3 \gamma}) Y(v^\gamma)$ computation and the $Y(v^{5 \gamma}) X(u^\gamma)$ version in \cite{Fleury:2017eph} are significantly different. Given the complexity of either computation, their agreement strongly vindicates the validity of the hexagon tiling approach to higher-point functions in ${\cal N} = 4$ SYM. To obtain an even more solid result it would be nice to run both schemes on all available tessellations, and for some correlation functions of half-BPS operators of higher length.

Next, our two-magnon computation as well as that of \cite{Fleury:2017eph} requires introducing a momentum factor $e^{\pm \frac{i}{2} p_2}$ when there is one scalar constituent at point 1, and $e^{\pm \frac{i}{2} p_1}$ for one scalar at point 2. Any rule one will assign must be compatible with earlier calculations, in particular with the original application of the gluing prescription in \cite{Basso:2015zoa}, so recovering the two-loop contribution to the structure constant for fusing two length two half-BPS operators into one twist two operator. However, that calculation was only done for fundamental magnons, obtaining a trace-like structure. From here the authors of \cite{Basso:2015zoa} conjectured that using bound states in the gluing one will always find parts of the transfer matrix. This transfer matrix proposal has been tested (also for the structure constant of two higher-length BPS with one twist 2 operator) to the next loop-order in \cite{Eden:2015ija,Basso:2015eqa} and is undoubtedly successful.

The problem of reconstructing the transfer matrix from the parts arising from the scattering of virtual magnons --- thus from bound state $S$-matrix elements --- has been analysed in \cite{Arutyunov:2009iq}. We illustrate in Appendix \ref{app:D} that the step becomes non-trivial for the structure constant computation from \cite{Basso:2015zoa}, because of the partitioning of the physical excitations into two groups. This enforces a special rule on how to insert transverse bound states on an edge, see Appendix \ref{app:D}. The re-constitution of the transfer matrix has to happen before the picture is changed from the string frame to the spin chain frame of the integrable system. As pointed out in \cite{Arutyunov:2009iq}, bound states $(\psi^1)^{a-k} (\psi^2)^k$ and $\phi^1 \phi^2 (\psi^1)^{a-1-k} (\psi^2)^{k-1}$ have to mix in an intricate fashion. Momentum dressing would have to arise in the moment of changing the picture to the spin chain frame, and could only cause a global factor multiplying the result of the mixing process. Therefore states with or without $\phi^1 \phi^2$ would automatically be treated in the same manner. Finally, as the states without scalar constituents do not receive momentum decorations, one would expect that the insertion of both scalars into one bound state is also neutral. 

With the present effort we confirm that the $J$-charge reacts like the internal symmetry charge $R$ in that it must have opposite eigenvalues on $\phi^1$ and $\phi^2$ constituents of bound states. Accordingly, matrix elements with and without $\phi^1 \phi^2$ can be added. The next question is if $J$  takes the same or the opposite eigenvalue as $R$ on, say, $\phi^1$. As in \cite{Fleury:2016ykk} we cannot avoid the conclusion that the one-magnon gluing necessitates an average over both choices.

As has been mentioned, a scalar bound state constituent introduces momentum dressing at the opposite end. W.r.t. the transfer matrix question, we can therefore not learn much about momentum dressing for the virtual magnon from an $sl(2)$ vacuum as in \cite{Basso:2015zoa}. In this respect it would be instructive to study three-point functions of two BPS and one operator with scalar excitations in an $su(2)$ sector.

The $\cX, \tilde \cX$ matrix elements cannot receive the same momentum decorations as $\cZ_{55}, \cZ_{66}$ in our $\gamma$ rotation as well as that of \cite{Fleury:2017eph}, so the momentum dressing is indeed sensitive to flavour. The considerations at the end of the last section show that link between $J$-charge of the bound states and the dressing by $e^{\pm \frac{i}{2} p}$ factors cannot be static: our calculation requires connecting $P_1 \, = \, e^{+\frac{i}{2} p_1}$ to $j_2$ (the $J$-charge at point 2) and $P_2$ to $j_1$, whereas in \cite{Fleury:2017eph} one has a rule of the same type, but with the association $1/P_1 \sim j_2$ and $P_2 \sim j_1$, a mismatch that we have not attempted to resolve. This may suggest that the momentum dressing ought to be assigned after mirror rotation to the physical edge on which the scattering takes place. 

Adopting this point of view, the $P$-dressing in Table 3 can be understood as a momentum operator acting to the left of a bound state with a single scalar in agreement with \cite{Arutyunov:2006yd} or as a $Z$-marker prescription as in Appendix A.1 of \cite{Fleury:2017eph}. To reproduce all entries in the table we need a $\oplus, \ominus$-grading, and the $J$-assignment would follow $P$; there is no link to $R$. We stress once again, that our two-magnon computation does not require an averaging over the momentum factors, though; one grading is sufficient.
  
Unfortunately, we did not fully achieve our aim of re-summing all residues in closed form. Nevertheless, we can analytically sum quite a few contributions. To this end we employed two techniques: first, using differentiation in the modulus to remove isolated factors $1/a^2, \, 1/a, \, 1/b^2, \, 1/b$ with the bound state lengths $a,b$, the coefficients in the series consist of $\Gamma$ functions with arguments built of only $a-k, \, b-l, \, k, \, l$. We can then shift the $a, \, b$ counters to decouple the sums. The resulting multiple geometric series can easily be summed into rational functions. Finally, integration in the modulus will make these into hyperlogarithms. Second, we sum by brute force, obtaining at some stage a hypergeometric function for which we immediately substitute its Euler integral representation. The integrand is then of the same type as before so that we can take the next sum. If another hypergeometric function is met, we substitute the kernel and so on. In a final step, hyperlogarithms are obtained from the parameter integrals.

Especially the second technique is quite versatile. The hypergeometric function has been used in attempts to sum up POPE \cite{Basso:2010in,Basso:2013vsa,Basso:2013aha,Basso:2014koa,Basso:2014nra} contributions \cite{Cordova:2016woh,Lam:2016rel}, and similar ideas should have applications also in related problems, e.g. in the exact evaluation of Mellin-Barnes representations of Feynman integrals \cite{Smirnov:1999gc}. On the other hand, there are order of limits issues, that render certain sequences of the operations impossible, while others allow us to get through to the end. This is reminiscent of the evaluation of ``linearly reducible'' integrals from a representation in terms of Feynman parameters \cite{Brown:2009ta,Panzer:2014caa,Panzer:2015ida}. Again, we should try to draw a link to those techniques.

The parts of our computation solvable by these methods point at a class of functions: the sum-integrals in our exercise are of the form rational function times weight two hyperlogarithms. The rational functions have a \emph{multilinear} polynomial denominator; the numerators consist of monomials that also occur in the respective denominators. For the pure functions we could pin down a basis with the same terms as letters in the symbol. The relative simplicity of the functions is certainly to some extent a one-loop effect, although explicit results for higher-loop BPS four-point functions \cite{Drummond:2013nda,Eden:2011we,Chicherin:2015edu} do not point at substantial complications higher up in the expansion. Naturally though, letters occurring in the denominators will at higher loop order appear in the symbol of the hyperlogarithms, too: by way of example, the integrals in the three-loop four-point functions of BPS operators \cite{Drummond:2013nda} can contain the letter $z-\bar z$ that we were happy to eliminate in the present exercise.

As we had pointed out, our first integration technique starts on a rational ``root'' (or several) to which polylogarithm levels are added by integration in the modulus of cross ratios. At this point it becomes important that the letters being integrated are multilinear in the four variables of the problem, whereby one can continue the process possibly indefinitely\footnote{We thank O. Schnetz for a discussion on this topic.}. It would be very interesting to study from a theoretical perspective which rational factors or letters in hyperlogarithms are generated from which starting point, e.g. to achieve an understanding of the function space for higher-loop corrections to the process studied here, or to understand more generally what kind of algebra such moves generate. In the amplitude literature, cluster algebras \cite{Golden:2013xva} have been singled out as a relevant algebraic structure. Our case presumably requires a more general concept due to the rational factors. For a start one could try to analyse the re-summation of the POPE for amplitudes from the perspective of our first integration scheme and try to establish a link to cluster algebras. A dream result of this type of programme would be a classification of function spaces for BPS correlators as for the six- and seven particle amplitudes 
\cite{Dixon:2011pw,Dixon:2013eka,Dixon:2014voa,Caron-Huot:2016owq,Golden:2014xqa,Dixon:2016nkn}. 

In the computations here presented, sum-integrals with the same dependence on the internal-space cross ratios should be added up (those related by swapping $(\psi^1)^{a-k} (\psi^2)^k$ for $\phi^1 \phi^2 (\psi^1)^{a-1-k} (\psi^2)^{k-1}$). The resulting expressions are single-valued \cite{Brown:2004Logs}. In the context of scattering amplitudes single-valuedness enables one to suppress half the variables; the full result can be reconstructed, see e.g. \cite{Broedel:2015nfp}. In our case the rational factors impede any quick success; to establish reconstructability would be very valuable.

Last, our results fall into pieces characterised by the denominator of the rational functions. It would be interesting to study what property of the scattering matrix is responsible for this --- can one reasonably organise the $S$-matrix elements in several simpler terms? Do the combinations of similar matrix elements simplify prior to integration and summation? Conversely, we have been able to push the evaluation of $S(\cX)$ much further than for the other matrix elements because there is an explicit writing of $\cX$. In Appendix \ref{app:A} a substantial simplification of the $\cY$ matrix elements is presented. On this basis we will try to push the calculation of the sum-integrals from the diagonal $\cY$ elements to the same level as the $\cX$ analysis presented in Appendix \ref{app:B}. Last, simplifications of the $\cZ$ elements will be presented in future work.

\section*{Acknowledgements}

D.~le Plat's work is funded by the Humboldt-Universit\"at zu Berlin within the Excellence Initiative of the states and the federal government. B.~Eden wishes to acknowlegde discussions with G.~Arutyunov, J.~Broedel, C.~Duhr, S.~Frolov and O.~Schnetz.
 MdL was supported by SFI, the Royal Society and the EPSRC for funding under grants UF160578, RGF$\backslash$EA$\backslash$181011, RGF$\backslash$EA$\backslash$180167 and 18/EPSRC/3590.
 A.S.'s work is funded by ETH Career Seed Grant No.~SEED-23 19-1, as well as by the NCCR SwissMAP, funded by the Swiss National Science Foundation.

\appendix

\section{The bound-state S-matrix}
\label{app:A}

In this appendix we list several identities and simplifications for the bound-state S-matrix which are useful for computing residues. We consider two bound states with bound state numbes $K_1,K_2$ and evaluation representation parameters $u_1,u_2$.

First, the central building block is the $\cX$-matrix which scatters fermionic states. This fundamental building block also appears in $\cY,\cZ$. From Yangian symmetry it can be shown that $\cX$ satisfies the following recursion relations
\begin{align}
    \cX^{k+1,l}_n &=\frac{1}{\bar{k}} \bigg[\frac{(\bar{n}-k) (\bar{k} + \bar{l}-\bar{n})}{\delta + \Sigma K-k-l-1} \cX^{k,l}_n + 
    \frac{(\bar{n}+1) (\delta+ \delta K +l-n+1)}{\delta + \Sigma K-k-l-1} \cX^{k,l}_{n-1}, \bigg] \label{eq:Xkplus}\\
    \cX^{k-1,l}_n &= \frac{1}{k} \bigg[\frac{(n-\bar{k}) (k+l-n)}{\delta - \Sigma K+k+l+1} \cX^{k,l}_n + \frac{n+1) (\delta - \delta K -l+n-1)}{\delta - \Sigma K+k+l+1} \cX^{k,l}_{n+1}, \bigg] \label{eq:Xkmin}\\
    \cX^{k,l+1}_n &= \frac{1}{\bar{l}}\bigg[ \frac{(\bar{k}+\bar{l}-\bar{n}) (\delta -\delta K -l+n)}{\delta +\Sigma K-k-l-1} \cX^{k,l}_n + \frac{(\bar{n}+1) (\bar{k}+\bar{l}-l-\bar{n}-1)}{\delta +\delta K-k-l-1} \cX^{k,l}_{n-1} \bigg] \label{eq:Xlplus}\\
    \cX^{k,l-1}_n &= \frac{1}{l}\bigg[ \frac{(k+l-n) (\delta + \delta K + l-n)}{\delta -\Sigma K+k+l+1} \cX^{k,l}_n + \frac{(n+1) (k+l-\bar{l}-n-1)}{\delta -\delta K+k+l+1} \cX^{k,l}_{n+1} \bigg]\label{eq:Xlmin},
\end{align}
where $\delta = u_1-u_2, 2\Sigma K = K_1+K_2,2\delta K = K_1-K_2$ and $\bar{k} = K_1 -k-1,\bar{l} = K_2-l-1,\bar{n} = K_1-n-1$. Notice that the cases $\cX^{k\pm1,l}_n$ and $\cX^{k,l\pm1}_n$ are each related by switching barred and unbarred indices. From these relations we see that for fixed $k,l$ all $\cX$-matrices with shifted indices can be brought into a standard form $\cX^{k,l}_n,\cX^{k,l}_{n\pm1}, \ldots$. 

Finally, by successively using \eqref{eq:Xkplus} and \eqref{eq:Xkmin} we obtain the following 
\begin{align}
    \cX^{k,l}_{n-1} =& \frac{(n+1) (k-\bar{l}-n-1) (\delta -\delta K-l+n+1)}{(\bar{n}+1) (k+l-n+1) (\delta +\delta K+l-n+1)}\cX^{k,l}_{n+1} + \nonumber\\
    &~\Bigg[\frac{(n+1) \bar{n} (\bar{l}-k-l+n)}{(\bar{n}+1) (k+l-n+1) (\delta +\delta K +l-n+1)} + \nonumber\\
    &\quad+\frac{(k-\bar{l}-n) (\delta +\delta K +l-n)}{(\bar{n}+1) (k+l-\bar{l}-n+1)}\\
&   \quad-\frac{(l+1) \bar{l} (\delta -\delta K+k+l+2) (\delta +\Sigma K-k-l -1)}{(\bar{n}+1)
   (k+l-n+1) (k+l-\bar{l}-n+1) (\delta +\delta K +l-n+1)}\Bigg]\cX^{k,l}_n \nonumber
\end{align}
From this we can also remove any $\cX$-matrix whose $n$ index is shifted by a negative integer. This can now be used to compare different, possibly equivalent ways of writing the other entries of the bound state S-matrix.

We can simplify the $\cY$-matrix as well. In particular, we need the diagonal elements, which can be written as
\begin{equation}
\begin{aligned}
\mathcal{Y}_{11} =&\, \frac{x^+_1-x^+_2}{x^-_1 -x^+_2}\frac{1}{U_1}
\bigg[
\frac{K_2-l}{K_2-k-l+n}\frac{\delta -\Sigma K+k+1}{\delta-\Sigma K+k+l+1}\frac{\delta+\delta K+l-n}{\delta+\delta K}\mathcal{X}^{k,l}_n+\\
&\quad\frac{K_1-k-1}{K_2-k-l+n}\frac{\delta-\delta K-l+n+1}{\delta-\Sigma K+k+l+1}\frac{l}{\delta +\delta K}\mathcal{X}^{k+1,l-1}_n
\bigg],\\
\mathcal{Y}_{22} =&\, \frac{x^-_1-x^-_2}{x^-_1-x^+_2} U_2
\bigg[
\frac{K_1-k}{K_1-n}\frac{\delta-\delta K-l+n}{\delta-\delta K}\frac{\delta-\Sigma K+l+1}{\delta-\Sigma K+k+l+1}\mathcal{X}^{k,l}_n+\\
&\quad \frac{(K_2-l-1)}{K_1-n}\frac{\delta+\delta K+l-n+1}{\delta-\delta K}\frac{k}{\delta-\Sigma K+k+l+1}\mathcal{X}^{k-1,l+1}_n
\bigg],\\
\mathcal{Y}_{33} =&\,  \frac{ x^-_1-x^-_2  }{x^+_1-x^-_2}U_1
\bigg[
\frac{\delta-\delta K+k+1}{\delta-\delta K}\frac{\delta+\delta K+l-n}{\delta -\Sigma K+k+l+1}\mathcal{X}^{k,l}_n-\\
&\quad \frac{K_1-k-1}{\delta-\delta K}\frac{\delta-\delta K-l+n+1}{\delta-\Sigma K+k+l+1}\mathcal{X}^{k+1,l-1}_n
\bigg],\\
\mathcal{Y}_{44} =&\,  \frac{x^+_2-x^+_1}{x^+_1-x^-_2} \frac{1}{U_2}
\bigg[
\frac{\delta+\delta K+l+1}{\delta+\delta K}\frac{\delta-\delta K-l+n}{\delta-\Sigma K+k+l+1}\mathcal{X}^{k,l}_n-\\
&\quad \frac{K_2-l-1}{\delta+\delta K}\frac{\delta+\delta K+l-n+1}{\delta-\Sigma K+k+l+1}\mathcal{X}^{k-1,l+1}_n
\bigg].
\end{aligned}
\end{equation}
We see that all elements of $\cY$ can be written as a linear combination of $\cX^{k\pm1, l\mp1}_n$ with some simple rational prefactors in $x^\pm$ and $u_i$. We can use the relations \eqref{eq:Xkplus}-\eqref{eq:Xlmin} to bring everything in the form of a similiar linear combinations of $\cX^{k,l}_{n\pm1}$ and $\cX^{k,l}_{n}$, but these expressions turn out to be less compact. In particular, in this form the prefactor of $\cX^{k,l}_{n}$ contains a large amount of simple rational function functions of $\delta$ which do not recombine to something simple. In each of these forms, the pole and residue structure is completely apparent and follows from the pole structure of $\cX$. 

Finally, we turn to $\cZ$. There are some useful relations that follow from the fermionic $\mathfrak{su}(2)$ symmetry and from a simple relabelling of states. We have for $i=1,2,3,4$ that
\begin{align}
&(\cZ^{k,l}_{n+1})_{i6} = - (\cZ^{k,l}_{n})_{i5},
&&(\cZ^{k+1,l-1}_{n})_{6i} = - (\cZ^{k,l}_{n})_{5i}, 
&&(\cZ^{k,l}_{n})_{65} = (\cZ^{k+1,l-1}_{n-1})_{56} .
\end{align}
For the diagonal elements we find
\begin{align}
 (\cZ^{k,l}_{n})_{55} &= \cX^{k,l-1}_n - (\cZ^{k,l}_{n+1})_{65}, \\
 (\cZ^{k,l}_{n})_{56} &= \cX^{k-1,l}_n - (\cZ^{k,l}_{n+1})_{66},\\
 (\cZ^{k,l}_{n})_{55} &= (\cZ^{k+1,l-1}_{n+1})_{66} ,
\end{align}
which shows that $\cZ_{55}$ and $\cZ_{66}$ basically have the same poles and residues after a relabelling of the indices.

From (6.11) of \cite{Arutyunov:2009mi} we can see that $\cZ$ can be written as a linear combination of rational factors multiplying various shifted versions of $\cX$. However, after using \eqref{eq:Xkplus}-\eqref{eq:Xlmin} it turns out that that can be reduced to just rational factors multiplying $\cX^{k,l}_{n\pm1}$. However, in this form, the prefactors are quite unwieldy but again, the pole structure becomes transparent. Finally, from (6.11) we also see that there is a potential problematic pole of the form $x^+_1x^+_2 - x^-_1x^-_2=0$ in $\cZ$. Fortunately, this pole appears to be spurious and is a consequence of how we decided to write $\cZ$ in (6.11). It seems to be absent when choosing a different presentation.

\section{$S_\mathrm{log}(\cX)$ and $S_\mathrm{mat}(\cX)$}
\label{app:B}

For scattering by the $\cX$-matrix the integral is
\begin{align}
\label{eq:integralX}
I_\cX \, = \, \sum_{a=1}^\infty\sum_{b=1}^\infty\sum_{k=0}^{a-1}\sum_{l=0}^{b-1} \int~\frac{du \, dv}{4 \, \pi^2}\mu\left(u,v\right) \, e^{i\frac{\pm p_1^{3\gamma} \mp \, p_2^\gamma}{2}} W_1\, W_2 \, \Sigma^{ab} \, \cX_k^{k,l} \, ,
\end{align}
where
\begin{align}
W_1&=z_1^{-iu+\frac{a}{2}-k-\frac{1\mp1}{2}}\bar{z}_1^{-iu-\frac{a}{2}+k+\frac{1\pm1}{2}}\alpha_1^{\frac{1\mp1}{2}}\bar{\alpha}_1^{-\frac{1\pm1}{2}} \, , \\
W_2&=z_2^{-iv+\frac{b}{2}-l-\frac{1\pm1}{2}}\bar{z_2}^{-iv-\frac{b}{2}+l+\frac{1\mp1}{2}}\alpha_2^{\frac{1\pm1}{2}}\bar{\alpha}_2^{-\frac{1\mp1}{2}} \, , \\
\Sigma^{a b}&=\frac{\gfunc{1+\frac{a}{2}+iu}\gfunc{1+\frac{b}{2}-iv}\gfunc{1+\frac{a+b}{2}-i(u-v)}}{\gfunc{1+\frac{a}{2}-iu}\gfunc{1+\frac{b}{2}+iv}\gfunc{1+\frac{a+b}{2}+i(u-v)}} \, \\
\mu(u,v)&=\frac{a \, b \, g^4}{\left(u^2+\frac{a^2}{4}\right)^2\left(v^2+\frac{b^2}{4}\right)^2} \, ,
\end{align}
and $\cX_k^{k,l}$ the $\cX$-matrix (\ref{defX}). Factors of the type $z^{\frac{1\pm1}{2}}$ are caused by the scalar constituents of the bound states scattered; first, at constant bound state-lenghts $a,b$ there is a spinor field less in either bound state. On the other hand, $R$- and $J$-charge are attributed to the scalars, and the $\pm$ in these exponents refer to the $\oplus,\ominus$ grading defined in (\ref{defJ}). Since such factors do not touch upon the summations, they will be omitted for the rest of this appendix, c.f. formulae (\ref{reDefW}).

The measure is of order $O(g^4)$. In $3\gamma,\gamma$ kinematics the momentum dressing $e^{i\frac{(p_1^{3\gamma}-p_2^\gamma)}{2}}$ is of order $O(g^{-2})$. The $\pm,\mp$ signs in the exponent of the momentum decoration in (\ref{eq:integralX}) are correlated in the concepts of gradings of \cite{Fleury:2017eph}, but we could allow them to be independent. The momentum factor can then be of order $g^0$ or $g^2$, too, both of which obviously do not contribute at one loop.

In $3\gamma,\gamma$ kinematics the factor $D$ in the $\cX$-matrix and the $e^{i\frac{(p_1^{3\gamma}-p_2^\gamma)}{2}}$ dressing take the following form:
\beq
D^{3\gamma,\gamma} \, = \, \frac{y^+-x^-}{x^+-y^-}\frac{\sqrt{x^+x^-y^+y^-}}{x^-y^+} \, , \qquad e^{i\frac{(p_1^{3\gamma}-p_2^\gamma)}{2}} \, = \, \frac{\sqrt{x^+x^-y^+y^-}}{g^2} \, .
\eeq
Together with the measure expressed that yields
\beq
\tilde{\mu} \, = \, \frac{a \, b \, g^2\left(v^+-u^-\right)}{\left(u^+-v^-\right)(u^-)^2 \, {u^+} \, {v^-} \, (v^+)^2} + O(g^2) \, , \label{tildeMu}
\eeq
cf.\ Table~\ref{tab:2}, where the denominator factor $(u^+ - v^-)$ was absorbed into $\cX$ as explained in the main text. 
\beq
\mu\left(u,v\right) \, e^{i\frac{p_1^{3\gamma}-p_2^\gamma}{2}} \,
W_1 \, W_2 \, \Sigma^{ab} \, \cX_k^{k,l} \, = \, \tilde{\mu} \, \Sigma^{ab} \, \tilde{W}_1 \, \tilde{W}_2 \, \hat{\cX}_k^{k,l},
\eeq
where
\beq
\tilde{W}_1 \, = \, z_1^{-iu+\frac{a}{2}-k}\bar{z}_1^{-iu-\frac{a}{2}+k} \, , \qquad 
\tilde{W}_2 \, = \, z_2^{-iv+\frac{b}{2}-l}\bar{z_2}^{-iv-\frac{b}{2}+l} \, , \qquad
\hat {\cX}^{k,l}_k \, = \, \frac{\cX_k^{k,l}}{D} \, . \label{reDefW}
\eeq
To do the integrations we invoke the residue theorem closing the contour over the upper half-plane for $u$ and the lower half-plane for $v$. The $\cX$-matrix has only poles in the lower half plane for $\delta \, = \, u-v$ so that no residue from poles of the matrix can contribute. Therefore, we can restrict to poles of the measure and the phase. As we have argued at the end of Section \ref{sec:1} we cannot localise both integrations using two poles from the phase. Hence we can choose both poles from the measure or one from the measure and one from the phase, as listed in equation (\ref{threeRes}). In this appendix we elaborate the first case, so $u \, = \, i \, a/2, \, v \, = \, - i \, b/2$. Finally, (\ref{tildeMu}) consists of two terms, one with a double pole in $(u^-)^2$ and one with a double pole $(v^+)^2$. We restrict our attention to the first one: $\cX$ describes the scattering of two bound states of equal type, whereby the other case can be retrieved by flip symmetry.

The double pole in $u$ creates a derivative on the remaining part of the integrand. To be sensitive to that, we introduce a shift of the integration variable $u \rar u + i \, \epsilon$. In this manner, taking the derivative can be left to the very end of the calculation. In practice, we simply work to first order in $\epsilon$.

The derivative could now act on the term $1/u^+ \rar -i(a + \epsilon)$. To simplify the calculation we will rather drop $\epsilon$ here; the first order correction $\epsilon \, i/a^2$ creating what we have called $S_\mathrm{mes}$ can be retrieved from the leading result $S_\mathrm{log}$ by integration in $|z_1|$ as described in Section \ref{sec:2}. Remarkably, these contributions are pure functions over the seventh and eighth denominator listed in (\ref{dens}).

So, without loss of generality
\begin{align}
\mathrm{res}_{\tilde{\mu}}\left(u=i \frac{a}{2}\right)=\frac{- i \, b \, g^2}{\left(ia+i\epsilon-v^-\right)v^+v^-}
\end{align}
and 
\begin{align}
\mathrm{res}_{\tilde{\mu}}\left(u=i \frac{a}{2},v=-i \frac{b}{2}\right)=\frac{g^2}{\left(a+b+\epsilon\right)}.
\end{align}
Henceforth we will also drop the overall $g^2$.

The phase and the matrix part have to be evaluated at the poles of the measure. The result is a Taylor series in ascending powers of $z_1,\bar{z}_1$, but descending powers of $z_2,\bar{z}_2$ with ratios coefficients made of $\Gamma$- functions from the phase and the matrix. The remaining denominator of the measure can be cancelled by shifting the argument of a $\Gamma$-function in the numerator of the phase. The $a,k$ and $b,l$ sums can be decoupled, respectively,  by swapping the order of summation and shifting indices as in Section \ref{sec:2}. Schematically, we find
\begin{align*}
\sum_{a=1}^\infty\sum_{b=1}^\infty\sum_{k=0}^\infty\sum_{l=0}^\infty\sum_{m=0}^ks\left(a+k,b+l,k,l,m\right),
\end{align*}
where the fifth sum over $m$ comes from the $\cX$-matrix (\ref{defX}) itself. We shift again to obtain an expression of the form
\begin{align*}
\sum_{b=1}^\infty\sum_{l=0}^\infty\sum_{k=0}^\infty\sum_{m=0}^\infty\sum_{a=1}^\infty s\left(a+k+m,b+l,k+m,l,m\right).
\end{align*}

In trying to fix a sequence of operations one would tend to pick one of the simplest sums at every stage. We observe here that the coefficients of the $a$ and $m$ sums involve two numerator and two denominator $\Gamma$-functions, the others are more complicated.
Summing first over $a$ gives a hypergeometric function:
\beq
\sum_{a=1}^\infty\frac{\gfunc{a+k}\gfunc{a+b+\epsilon}z_1^a}{\gfunc{a}\gfunc{1+a+k+\epsilon}} \, = \, \frac{z_1\gfunc{1+k}\gfunc{1+b+\epsilon}}{\gfunc{2+k+\epsilon}} {}_2F_1\left(1 + k, 1 + b + \epsilon, 2 + k + \epsilon, z_1\right) \, .
\eeq
Replacing the hypergeometric sum by its integral representation
\beq
\label{eq:2F1Int}
\begin{aligned}
_2F_1\left(a,b;c;x\right) & = \, \frac{\gfunc{c}}{\gfunc{b}\gfunc{c-b}}\int\limits_0^1\mathrm{d}s~\frac{s^{b-1}\left(1-s\right)^{c-b-1}}{\left(1-x s\right)^a} \\
& = \, \frac{\gfunc{c}}{\gfunc{a}\gfunc{c-a}}\int\limits_0^1\mathrm{d}s~\frac{s^{a-1}\left(1-s\right)^{c-a-1}}{\left(1-x s\right)^b}
\end{aligned}
\eeq
we find more complicated powers $(1-s)^n, \, (1 - s \, x)^n$ but the integrand is of the same type as before, because all coefficients are $\Gamma$-functions. The strategy is now to leave the integrations to the end and to try and take the next sum. Some experiments about the sequence of operations are needed; there are several paths through the calculation. In this respect the situation is similar to that of executing parametric integrals in solving Feynman-integrals by linear reducibility \cite{Brown:2009ta}. We choose to sum in $m, \, k, \, l$ next. Each of these sums is then of the $_2F_1$ type and necessitates introducing an integration parameter. We call the four parameters $s, \, t, \, r, \, u$, respectively. The procedure thus transforms four sums into integrals at comparative ease. Changing the sequence one may fall upon higher hypergeometric functions at intermediate stages, which would yield more parameters. As a rule, of the two choices in (\ref{eq:2F1Int}) one should choose the one that does not introduce divergences into the integral. This is not always possible; it may also be convenient to make a divergent choice to simplify subsequent steps.

The last sum over $b$ cannot be taken (or, in other sequences of operations, the last remaining sum) because the factors of $\gfunc{1-b}$ and $\gfunc{-\epsilon}$ appear in the denominator. For once, for every integer value of $b$ the summand is identically zero. On the other hand,  $\gfunc{-\epsilon}$ signals a divergence in some parametric integral. We might now hope that the two effects compensate each other. However, sum and integral do not commute in such a situation; summing first (in some regularisation) we lose the divergence of the integral and find zero. Therefore, a divergent integral has to be done before taking the sum in $b$.

Normally, executing the last integration first is not helpful because one will simply re-instate the hypergeometric function that had been written as an integral. Nonetheless, in this case it helps to deal with the $u$-integral first in an expansion with respect to $\epsilon$. Partially fractioning the integrand isolates the divergence:
\begin{align}
\label{eq:uint}
&\frac{\left(1-u\right)^{-\epsilon}u^{b+\epsilon}}{\left(1-u\right)\left(1-t \, \bar{z}_1+\frac{1}{\bar{z}_2}u\left[r  \, s \, t \, \bar{z}_1-1\right]\right)}\\
=&\frac{\left(1-u\right)^{-1-\epsilon}u^{b+\epsilon}}{\left(1-t \, \bar{z}_1-\frac{1}{\bar{z}_2}+\frac{\bar{z}_1}{\bar{z}_2} r \, s \, t \right)}+\frac{\left(1-u\right)^{-\epsilon}u^{b+\epsilon}}{\left(1-t \, \bar{z}_1-\frac{1}{\bar{z}_2}+\frac{\bar{z}_1}{\bar{z}_2} \, r \, s \, t\right) \left(1-t \, \bar{z}_1+\frac{1}{\bar{z}_2} \, u \left[r \, s \, t \, \bar{z}_1-1\right]\right)}\notag
\end{align}
In the first term the divergence of the integral at $ u \, = \, 1$ leads to a factor of $\gfunc{-\epsilon}$ which cancels against the denominator. Therefore this term is non-vanishing for $\epsilon \rar 0$. The integration of the second part would re-create a hypergeometric function. Yet, it is not divergent at the end points of the integration range $u \, \in \, [0,1]$. The pole in the second term can be moved away from the unit interval e.g. by assuming the $\bar z_1, \bar z_2$ complex, which is the case in Euclidean kinematics at any rate. We may therefore expand this term in $\epsilon$ to first order. 

The integration over $r$ is now elementary in both terms. It yields a factor of $\gfunc{1-b}$ which cancels against the denominator. Now, the sum over $b$ can be taken. For the first summand of \eqref{eq:uint} we obtain
\begin{align}
-\frac{\epsilon \, z_2 (1-s)^\epsilon (1-t)^{\epsilon-1} z_1^{\epsilon+1} \bar{z}_1^\epsilon (1-s \, z_1)^{-\epsilon-1} (1-t \, \bar{z}_1)^{-\epsilon} \left(-1+ {}_2F_1\left[\epsilon+1,\epsilon+1;\epsilon;\eta\right)\right]}{s \, t \, \bar{z}_1-t \, z_2 \bar{z}_1+z_2-1} \, , \label{doubleDen}
\end{align}
where
\begin{align}
\eta \, = \, \frac{(t-1) (-t \, \bar{z}_1
   z_2+z_2+s \, t \, \bar{z}_1-1)}{(s \, z_1-1) (t \, \bar{z}_1-1) (z_2 (t \, \bar{z}_1-1)+1) \bar{z}_2}. 
\end{align}
Here, the hypergeometric function can be expressed in terms of logarithms as following \cite{Huber:2005yg,Huber:2007dx}:
\begin{align}
_2F_1\left(\epsilon+1,\epsilon+1;\epsilon;\eta\right)=\frac{\epsilon+\eta-\epsilon \, \eta \log\left(1-\eta\right)}{\epsilon\left(1-\eta\right)^2} + O(\epsilon^2) \label{he2F1}
\end{align}
In this term we have the integrations over $s,t$ left. Due to the double denominator $(1-\eta)^2$ in (\ref{doubleDen}) the $s$-integral, say, does not necessarily add a logarithm level. We find a rational term, whose $t$-integral yields the r.h.s of formula (\ref{sLogX}) for $S_\mathrm{log}(\cX)$. In the way explained in Section \ref{sec:2} we can obtain the $S_\mathrm{mes}$ contributions from here by integrating in the modulus of $z_1$ and $z_2$, respectively. 

Where the $s$-integral does create a logarithm, or else in the terms from $\log(1-\eta)$ from the numerator of (\ref{he2F1}), the $t$-integral yields hyperlogarithms of weight two. Further, we have to integrate the second part of \eqref{eq:uint} in the three parameters $s, \, t, \, u$. Here, the $s$-integral, say, has an integrand of the type $\mathrm{polynomial}/\mathrm{polynomial}^2$ and remains rational. Each of the remaining two integrations adds one  logarithm level. The full weight two result expressed in our basis of functions reads:
\begin{align}
S_\mathrm{mat,u}(X)&=\frac{y \, z \, L_X}{4\left(y \, z - a \, y \, z - b \,  y \, z - a \, b + a \, b \, y \ + a \, b \, z \right)}\notag\\
L_X&=-2 \, \text{Li}_2\left(\frac{a \, z -a+y-b \, }{(1-z) (1-a-b)}\right)+2 \, 
   \text{Li}_2\left(\frac{z}{1-y}\right)-2 \, \text{Li}_2(z)+2 \, \text{Li}_2(b)\notag\\
&-2 \, \text{Li}_2\left(\frac{a \, y \, z-y \, z-a \, b \, z+b \,  y \, z+a \, b-a \, b \, y}{(1-a) (1-b)(1-y-z)}\right)\notag\\
&+2 \, \text{Li}_2\left(\frac{y-a}{1-a}\right)-2 \, \text{Li}_2\left(\frac{a}{1-b}\right)-2 \, 
   \text{Li}_2\left(\frac{b}{1-a}\right)+2 \, \text{Li}_2(a)\notag\\
&-2 \, \log (1-z) \log (1-a-b)+2 \, \log (1-a) \, \log (1-z)\notag\\
&-2 \, \log ^2(1-a-b)-2 \, \log (1-a) \, \log
   (1-b) \\
&+4 \, \log (1-a) \, \log (1-a-b)+4 \, \log(1-b) \, \log (1-a-b)\notag\\
&-2 \, \log ^2(1-a)-2 \, \log(1-b) \, \log (1-y-z)-\log ^2(1-y-z)\notag\\
&-2 \, \log (1-y) \, \log (1-y-z)+2 \, \log(1-z) \log (1-y-z)\notag\\
&+3 \, \log^2(1-y)+2 \, \log(1-z) \, \log(1-b)-\log^2(1-z)-2 \, \log
   ^2(1-b)\notag\\
&+2 \, \log (1-a-b) \log (1-y-z)-2 \, \log (1-a) \log
   (1-y-z)\notag.
\end{align}

\section{$S(\cZ_{11})$}
\label{app:C}

In this appendix we present a check on the analysis in Section \ref{subsec:2:2}: instead of the differentiation/integration trick we use the method of Appendix \ref{app:B}. 

First, since the $\cZ_{11}$ element has the summation range $\sum_{a=1}^\infty \sum_{k = 0}^a$ and similarly for $b,l$ we have to split the sums as in \eqref{splitSum} at both ends when shifting the counters, as in the first computation. This introduces a number of special cases, which we do not discuss here, although they cannot be omitted, of course. To illustrate the method, suffice it to study
\begin{eqnarray}
\tilde S_1 & = & \sum_{a,b,k,l \, = \, 1}^\infty a_2^l \, b_1^k \, y_2^b \, z_1^a \frac{k^2 \, \Gamma[a + b] \Gamma[k + l]}{(a + k)^2 \Gamma[1 + a] \Gamma[1 + b] \Gamma[1 + k] \Gamma[1 + l]} \, , \\
\tilde S_2 & = & - 2 \sum_{a,b,k,l \, = \, 1}^\infty a_2^l \, b_1^k \, y_2^b \, z_1^a \frac{\Gamma[a + b] \Gamma[k + l]}{(a + k)(b+l) \Gamma[1 + a] \Gamma[1 + b] \Gamma[1 + k] \Gamma[1 + l]} \, .
\end{eqnarray}
We can find the analogous sum $\tilde S_3$ by flip symmetry. Due to the factors $(a+k),(b+l)$ (before shifting the counters $a,b$), holomorphic and the antiholomorphic part do not decouple anymore so that multiple hypergeometric series arise.

Let us focus on $\tilde S_1$ first. The $b$ and $l$ integrals are still simple geometric series. We find
\beq
\tilde S_1 \, = \, \sum_{a,k \, = \, 1}^\infty b_1^k (1 - (1-a_2)^{-k}) \, z_1^a (1 - (1-y_2)^{-a}) \, \left( -\frac{1}{(a+k)^2} + \frac{1}{a (a+k)} \right) \label{S1bits}
\eeq
with the two types of sums
\beq
\begin{aligned}
\Sigma_1 & = \, \sum_{a,k \, = \, 1}^\infty \frac{u^a \, v^k}{(a+k)^2} \, = \, \sum_{a=1}^\infty \frac{u^a \, v}{(1+a)^2} \, {}_3F_2[\{1,1+a,1+a\},\{2+a,2+a\},v] \\
& = \, \int_0^1 ds \, dt \, v \sum_{a \, = \, 1}^\infty u^a (1 - s)^a t^a (1 - s \, t \, v)^{-(1 + a)}  \\
& = \, \int_0^1 ds \, dt \frac{(1-s) \, t \, u \, v}{(1 - s \, t \, v) (1 - t \, u + s \, t \, u - s \, t \, v)} \, = \, \frac{v \, \mathrm{Li}_2(u) - u \, \mathrm{Li}_2(v)}{u-v} \, , \\
\end{aligned}
\eeq
\beq
\begin{aligned}
\Sigma_2 & = \, \sum_{a,k \, = \, 1}^\infty \frac{u^a \, v^k}{a(a+k)} \, = \, \sum_{a=1}^\infty \frac{u^a \, v}{a} \, {}_2F_1[1,1+a,2+a,v] \\
& = \, \int_0^1 ds \, v \, \sum_{a \, = \, 1}^\infty \frac{u^a (1-s)^a (1- s \, v)^{-(1+a)}}{a} \, = \, - \int_0^1 ds \frac{v \log\left[ \frac{1-u+s \, u - s \, v}{1 - s \, v} \right] }{1- s \, v} \\
& = \, \mathrm{Li}_2\left[ \frac{u-v}{u(1-v)} \right] - \mathrm{Li}_2\left[\frac{u-v}{u} \right] + \log[1-v] \left( \frac{1}{2} \log[1-v] - \log[v] + \log[u] \right)
\end{aligned}
\eeq
The final result for $\tilde S_1$ from substituting the parts of \eqref{S1bits} is a little unwieldy so that we refrain from presenting it here. In the  $\tilde S_2$ sum we may start by summing over $k$, obtaining a $_2F_1$:
\begin{eqnarray}
\tilde S_2 & = & - 2 \, b_1 \sum_{a,b,l \, = \, 1}^\infty a_2^l \, y_2^b \, z_1^a \, \frac{ l \, \Gamma[a + b]}{(1 + a) (b + l) \Gamma[1 + a] \Gamma[1 + b]} \, {}_2F_1[1 + a, 1 + l, 2 + a, 
   b_1] \nonumber \\
   & = & - 2 \, b_1 \int_0^1 ds \, \sum_{a,b,l \, = \, 1}^\infty a_2^l \, s^a (1 - b_1 \, s)^{-(1+l)} y_2^b \, z_1^a \frac{l \, \Gamma[a + b]}{(b + l) \Gamma[1 + a] \Gamma[1 + b]} \\
   & = & 2 \, b_1 \int_0^1 ds \sum_{b,l \, = \, 1}^\infty a_2^l (1 - b_1 \, s)^{-(1+ l)} y_2^b (1 - (1 - s \, z_1)^{-b}) \frac{l}{b(b+l)} \nonumber \\
   & = & 2 \, a_2 \, b_1 \int_0^1 ds  \frac{1}{(1 - b_1 \, s)^2} \sum_{b \, = \, 1}^\infty  y_2^b (1 - (1 - s \, z_1)^{-b} )\frac{1}{b(1+b)} {}_2F_1\left[2, 1 + b, 2 + b, \frac{a_2}{1 - b_1 \,  s}\right] \nonumber
\end{eqnarray}
\begin{eqnarray}
\phantom{\tilde S_2}
   & = & 2 \, a_2 \, b_1 \int_0^1 ds \, dt \, \sum_{b \, = \, 1}^\infty (1 - b_1 \, s)^{b-1} (1 - t)^{b-1} t (1 - b_1 \, s - a_2 \, t)^{-(1+b)} y_2^b (1 - (1 - s \, z_1)^{-b}) \nonumber \\
   & = & - 2 \, a_2 \, b_1 \, y_2 \, z_1 \int_0^1 ds \, dt \, \frac{s \, t}{ 1 - b_1 \, s - a_2 \, t - y_2 + b_1 \, y_2 \, s + y_2 \, t - b_1 \, y_2 \,  s \, t}  \nonumber * \\
   && \qquad * \, \frac{1}{1 - b_1 \, s - a_2 \, t - y_2 + b_1 \, y_2 \, s + y_2 \, t - b_1 \, y_2 \, s \, t - z_1 \, s + b_1 \, z_1 s^2 + a_2 \, z_1 \, s \, t} \nonumber
 \end{eqnarray}
Here we can do the $t$-integral without creating $\log(\sqrt{\ldots})$. Upon putting $t \, = \, 0,1$ from the boundaries of the integration domain, the $s$-dependence separates into linear factors. Finally, integrating in $s$ we fall upon the usual function space. Putting together the explicit results for $\tilde S_1, \, \tilde S_2, \, \tilde S_3$, and the boundary terms we exactly reproduce \eqref{sZ11}, \eqref{zComp}.

\section{Towards the transfer matrix}
\label{app:D}

Consider the structure constant for fusing two protected operators $\mathcal{O}_2 \, = \, \mathrm{Tr}(\hat Z^2)$ (here $\hat Z$ is the co-moving vacuum of \cite{Drukker:2009sf,Basso:2015zoa}) into the twist two operator $\mathcal{O}_1 \, = \, \mathrm{Tr}(D^2 \hat Z^2)$ (schematically) in the Konishi multiplet, so the lowest non-trivial case of the series of structure constants studied in \cite{Basso:2015zoa}. At leading and subleading order in the coupling constants this is given by the asymptotic result. Wrapping corrections come in the ``opposite channel'' first \cite{Basso:2015zoa}, so if the non-protected operator is at the $0 \gamma$ physical edge, there is a virtual magnon on the virtual edge opposite to it. It will be moved to the physical edge by a $-3 \gamma$ rotation to line up with the physical excitations on both hexagons. In \cite{Basso:2015zoa} the calculation is done for virtual magnons of bound state length one only yielding the corresponding term of a transfer matrix. gluing on the opposite edge comes in at $O(g^4)$ because all edges have width one. The different mirror rotation for gluing over the virtual edges adjacent to the twist two operator leads to a leading order cancellation in the transfer matrix at bound state length one whereby such contributions are postponed to $O(g^6)$. It is then assumed that gluing by virtual magnons of higher bound state length will indeed produce the higher terms of the transfer matrix. Yet, this is not obvious due to a mixing effect described in \cite{Arutyunov:2009iq}. In this appendix we study how that mechanism could work despite of the partitioning of the Bethe roots built into the hexagon approach.

The integrand for the opposite wrapping configuration is of the familiar form \cite{Basso:2015zoa}
\beq
\label{int3gamma}
\begin{aligned}
& \mathrm{int}^{3\gamma}_b (u,\{u_1,u_2\})\\
& =\, \sum_{\alpha \cup \bar \alpha \, = \, \{u_1,u_2\}} (-1)^{|\bar{\alpha}|}\prod_{j \in \bar{\alpha}} e^{i p_j \ell} \prod_{\substack{i>j \\ j \in \bar{\alpha}, i \in \alpha}} S_{ji} \sum_{\mathcal{X}_b}(-1)^{f_{\mathcal{X}_b}} \braket{\mathcal{H}| \alpha,\mathcal{X}_b^{-3\gamma}} \braket{\mathcal{H}| \bar{\alpha},\tilde{\mathcal{X}}_b^{-3\gamma}}  
\end{aligned}
\end{equation}
with $\ell \, = \, 1$.
The splitting factor $e^{i p_j}$ and the S-matrix $S_{jk}$ capture the cutting of the spin chain. The physical particles have bound state length one. The mirror particle is a bound state $\mathcal{X}_b$ of length $b$ and is inserted on the front and the back hexagon. Rotating by 
$-3\gamma$ 
the bound states are on the right of the partitions of physical particles $\alpha$ and $\bar{\alpha}$. Evaluating the hexagon form factor $\mathcal{H}$ yields elements of the bound state S-matrix \cite{Arutyunov:2009mi} and the scalar $h$-factor \cite{Basso:2015zoa}. As for the fundamental magnons, a pair of entangled bound states $\mathcal{X}$ and $\tilde{\mathcal{X}}$ is inserted on the edges of the two adjacent hexagons, respectively, as can be seen from \eqref{int3gamma}. The sum is taken over all possible bound states.
Here, we propose the following conjugation rule for the insertion of bound states
\begin{align}
\mathcal{X}_b &= \ket{\dot{\phi}_1^{\gamma_1} \dot{\phi}_2^{\gamma_2} \dot{\psi}_1^{c_1} \dot{\psi}_2^{c_2} } \otimes \ket{ \phi_1^{\alpha_1} \phi_2^{\alpha_2} \psi_1^{a_1} \psi_2^{a_2}} \nonumber \\
\tilde{\mathcal{X}}_b &= \ket{\dot{\phi}_1^{\alpha_2} \dot{\phi}_2^{\alpha_1} \dot{\psi}_1^{a_2} \dot{\psi}_2^{a_1} } \otimes \ket{ \phi_1^{\gamma_2} \phi_2^{\gamma_1} \psi_1^{c_2} \psi_2^{c_1}}, \label{ConjugationRule}
\end{align}
with $\alpha_{1,2},\gamma_{1,2}=0,1$ and the bound state length $b=\alpha_1+\alpha_2+a_1+a_2=\gamma_1+\gamma_2+c_1+c_2$. The conjugation exchanges the left and the right part of the magnon and additionally the fields indices $1 \leftrightarrow 2$. In particular, for ``longitudinal'' bound states one has $\alpha_1 \, = \, \gamma_2, \alpha_2 \, = \, \gamma_1, a_1 \, = \, c_2, \, a_2 \, = \, c_1$ so that $\chi_b \, = \, \tilde \chi_b$. This rule also has a very nice interpretation in the sense of inserting the identity operator $\ket{\mathcal{X}}\bra{\tilde{\mathcal{X}}}$ on the virtual edge, where the scalar product is induced by the hexagon contraction. Our conjugation rule is necessary to obtain transfer matrix terms as in \cite{Basso:2015zoa,Eden:2015ija,Basso:2015eqa} also for bound states with length $b \, > \, 1$.

The notation used in the following will closely follow the one used in \cite{Arutyunov:2009iq}.
Let us firstly consider both physical excitations on the same hexagon. A hexagon containing only the bound state is only non-zero if the bound state is longitudinal, i.e. if the left part can be contracted on the right part.
There are four longitudinal bound states of length $b$:
\begin{align}
 \ket{\dot{\psi}_1^{l} \dot{\psi}_2^{b-l} } &\otimes \ket{\psi_1^{b-l} \psi_2^{l}}, \nonumber \\
 \ket{\dot{\phi}_2 \dot{\psi}_1^{l} \dot{\psi}_2^{b-l-1} } &\otimes \ket{ \phi_1 \psi_1^{b-l-1} \psi_2^{l}}, \nonumber \\
 \ket{\dot{\phi}_1 \dot{\psi}_1^{l} \dot{\psi}_2^{b-l-1} } &\otimes \ket{ \phi_2 \psi_1^{b-l-1} \psi_2^{l}}, \nonumber \\
 \ket{\dot{\phi}_1 \dot{\phi}_2 \dot{\psi}_1^{l-1} \dot{\psi}_2^{b-l-1} } &\otimes \ket{ \phi_1 \phi_2 \psi_1^{b-l-1} \psi_2^{l-1}}
\end{align}
By the Yang-Baxter equation we can scatter the physical magnons over each other first. Due to the contraction rule the scattering over the bound state then has to be diagonal w.r.t. the $l$ counter. As was observed in \cite{Arutyunov:2009iq}, the flavour of the bound state (i.e. $(\alpha_1,\alpha_2) \, = \, (1,0)$ or $(0,1)$) can oscillate, though, while it moves through the chain of physical magnons. Indeed this mixing is essential to build up the transfer matrix.
The relevant $S$-matrix elements will be 
$\mathcal{Y}^{0,l;2}_{0;2}$, $\mathcal{Z}^{0,l;1}_{0;1}$, $\mathcal{Z}^{0,l;2}_{0;2}$, $\mathcal{Z}^{0,l;1}_{0;2}$, $\mathcal{Z}^{0,l;2}_{0;1}$.
For the hexagon form factors the following amplitudes contribute
\begin{align}
\sum_{\mathcal{X}} \braket{\mathcal{H}| D_1D_2 \mathcal{X}^{-3\gamma}} &\braket{\mathcal{H}| \tilde{\mathcal{X}}^{-3\gamma}}= \braket{\mathcal{H}|D_1D_2} h(u_1,u^{-3\gamma}) \,  h(u_2,u^{-3\gamma}) \, * \nonumber \\
& \bigg[ -2  \mathcal{Y}^{0,l;2}_{0;1}(u_1,u^{-3\gamma}) \mathcal{Y}^{0,l;1}_{0;1}(u_2,u^{-3\gamma}) \\
&+ \mathcal{Z}^{0,l;1}_{0;1}(u_1,u^{-3\gamma}) \mathcal{Z}^{0,l;1}_{0;1}(u_2,u^{-3\gamma}) + \mathcal{Z}^{0,l;2}_{0;1}(u_1,u^{-3\gamma}) \mathcal{Z}^{0,l;1}_{0;2}(u_2,u^{-3\gamma})  \nonumber \\
&+ \mathcal{Z}^{0,l;1}_{0;2}(u_1,u^{-3\gamma}) \mathcal{Z}^{0,l;2}_{0;1}(u_2,u^{-3\gamma}) + \mathcal{Z}^{0,l;2}_{0;2}(u_1,u^{-3\gamma}) \mathcal{Z}^{0,l;2}_{0;2}(u_2,u^{-3\gamma}) \bigg] \nonumber \label{PartitionJoint}
\end{align}
The expression written in square brackets on the right hand side of eq. \eqref{PartitionJoint} is the bound state transfer matrix $\mathcal{T}(u)$ written in \cite{Arutyunov:2009iq}. 

The transfer matrix should arise independently of the partition of the physical particles $\alpha,\bar{\alpha}$ in eq. \eqref{int3gamma}. Therefore consider now one derivative on the front and the second on the rear hexagon. Using only longitudinal bound states it is not possible to create the matrix elements 
$\mathcal{Z}_{0;1}^{0,l;2}$ and $\mathcal{Z}_{0;2}^{0,l;1}$.
Therefore one has to include transverse bound states. Two of these contribute at length $b$ in this configuration:
\begin{align}
 \ket{\dot{\phi}_1 \dot{\phi}_2 \dot{\psi}_1^{l-1} \dot{\psi}_2^{b-l-1} } &\otimes \ket{ \psi_1^{b-l} \psi_2^{l}}, \nonumber \\
  \ket{\dot{\psi}_1^{l} \dot{\psi}_2^{b-l} } &\otimes \ket{ \phi_1 \phi_2 \psi_1^{b-l-1} \psi_2^{l-1}}
\end{align}
Conjugating the bound state on the second hexagon as stated in \eqref{ConjugationRule} and summing over all bound states yields the same matrix elements as eq. \eqref{PartitionJoint}.

As we had mentioned, the conjugation rule acts as an identity on longitudinal bound states. Since for bound state length $b \, = \, 1$ only longitudinal states appear, the same fundamental magnons $Y,\bar{Y},D,\bar{D}$ are inserted on both hexagons. However, inserting identical bound states with $b>1$ would not lead to the combination of matrix elements 
$\mathcal{Z}^{0,l;1}_{0;2} \mathcal{Z}^{0,l;2}_{0;1}$ and $\mathcal{Z}^{0,l;2}_{0;1} \mathcal{Z}^{0,l;1}_{0;2}$ but rather yield $\mathcal{Z}^{0,l;1}_{0;2} \mathcal{Z}^{0,l;1}_{0;2}$ and $\mathcal{Z}^{0,l;2}_{0;1} \mathcal{Z}^{0,l;2}_{0;1}$.

Next, one might try inserting $\bar{\mathcal{X}}$ on the second hexagon, where the bar means the exchange of the left and right part of the bound state. This exchanges $l \leftrightarrow b-l$ in the matrix elements from the second hexagon: $\mathcal{Z}^{0,l;1}_{0;2} \mathcal{Z}^{0,b-l;2}_{0;1}$ and $\mathcal{Z}^{0,l;2}_{0;1} \mathcal{Z}^{0,b-l;1}_{0;2}$. Luckily these have the symmetries $\mathcal{Z}^{0,b-l;1}_{0;2}=\mathcal{Z}^{0,l;1}_{0;2}$, $\mathcal{Z}^{0,b-l;2}_{0;1}=\mathcal{Z}^{0,l;2}_{0;1}$. However, the exchange $l \rar b-l$ also happens for the diagonal elements on the rear hexagon stemming from the longitudinal bound states, $\mathcal{Z}^{0,l;1}_{0;1}, \, \mathcal{Z}^{0,l;2}_{0;2}$. For those there is no symmetry of the aforementioned type.

Moreover, the appearance of transversal bound states allows us to draw conclusions on the dressing of $\ket{ \phi_1 \phi_2 \psi_1^{b-l-1} \psi_2^{l-1}}$. The mapping from the string frame to the spin chain frame introduces dressing factors on the states and the matrix elements. However, the dressing factor of the result does only depend on the final state according to \cite{Arutyunov:2006yd}. In this manner, scattering the longitudinal state $ \ket{\dot{\psi}_1^{l} \dot{\psi}_2^{b-l} } \otimes \ket{\psi_1^{b-l} \psi_2^{l}}$ contributes $\mathcal{Z}^{0,l;1}_{0;1}(u_1,u^{-3\gamma}) \mathcal{Z}^{0,l;1}_{0;1}(u_2,u^{-3\gamma}) + \mathcal{Z}^{0,l;2}_{0;1}(u_1,u^{-3\gamma}) \mathcal{Z}^{0,l;1}_{0;2}(u_2,u^{-3\gamma})$ to eq. \eqref{PartitionJoint}. Since the left chain contains only fermions after the scattering no dressing is needed.

Using the transversal bound states the element $\mathcal{Z}^{0,l;2}_{0;1}(u_1,u^{-3\gamma})$ scatters $\ket{\psi_1^{b-l} \psi_2^{l}}$ to $\ket{ \phi_1 \phi_2 \psi_1^{b-l-1} \psi_2^{l-1}}$. The final state does contain bosons and therefore a dressing factor might appear. At the same time, on the second hexagon $\ket{ \phi_1 \phi_2 \psi_1^{b-l-1} \psi_2^{l-1}}$ is scattered to $\ket{\psi_1^{b-l} \psi_2^{l}}$ giving rise to $\mathcal{Z}^{0,l;1}_{0;2}(u_2,u^{-3\gamma})$. Since the final state only contain fermions, no dressing is expected.
Putting the hexagons back together, one might thus obtain a dressing factor, leading to a dressing of the product $\mathcal{Z}^{0,l;2}_{0;1}(u_1,u^{-3\gamma}) \mathcal{Z}^{0,l;1}_{0;2}(u_2,u^{-3\gamma})$. Then again, there is no dressing factor using longitudinal bound states only, so we conclude that there is no dressing for $\ket{ \phi_1 \phi_2 \psi_1^{b-l-1} \psi_2^{l-1}}$. The mixing of the bound states $\ket{\psi_1^{b-l} \psi_2^{l}}$ and $\ket{ \phi_1 \phi_2 \psi_1^{b-l-1} \psi_2^{l-1}}$ was originally pointed out in \cite{Arutyunov:2009iq}.

We explicitly evaluated the bound state transfer matrices for up to six physical particles in the $0\gamma,-3\gamma$ kinematics. This also enlarges the number of possible combinations of the above $\mathcal{Z}$-matrix elements. We did indeed find in the cases considered that the numerator of the integrand can be represented in terms of Baxter polynomials $Q(u)=\prod_{i}(u-u_i)$ \cite{Basso:2015zoa,Eden:2015ija,Basso:2015eqa} as 
\begin{equation}
T_b^{(0)} (u^\gamma) \, \propto \, Q(u^{[b+1]})+Q(u^{[-b-1]})-Q(u^{[b-1]})-Q(u^{[-b+1]})
\end{equation}
when the zero momentum/level matching constraint $\prod_j e^{i p_j} \, = \, 1$ is fullfilled, as required for the structure constant computation \cite{Basso:2015zoa}.

\bibliographystyle{JHEP}
\bibliography{refs}

\makeatletter \@ifundefined{Sphere}{\newcommand{\Sphere}{\text{S}}}{}
  \@ifundefined{AdS}{\newcommand{\AdS}{\text{AdS}}}{}
  \@ifundefined{CFT}{\newcommand{\CFT}{\text{CFT}}}{}
  \@ifundefined{CP}{\newcommand{\CP}{\text{CP}}}{}
  \@ifundefined{Torus}{\newcommand{\Torus}{\text{T}}}{}
  \@ifundefined{superN}{\newcommand{\superN}{\mathcal{N}}}{}
  \@ifundefined{grpOSp}{\newcommand{\grpOSp}{\mathrm{OSp}}}{}
  \@ifundefined{grpPSU}{\newcommand{\grpPSU}{\mathrm{PSU}}}{}
  \@ifundefined{grpSU}{\newcommand{\grpSU}{\mathrm{SU}}}{}
  \@ifundefined{grpU}{\newcommand{\grpU}{\mathrm{U}}}{}
  \@ifundefined{grpD}{\newcommand{\grpD}{\mathrm{D}}}{}
  \@ifundefined{grpSL}{\newcommand{\grpSL}{\mathrm{SL}}}{}
  \@ifundefined{grpSp}{\newcommand{\grpSp}{\mathrm{Sp}}}{}
  \@ifundefined{grpUSp}{\newcommand{\grpUSp}{\mathrm{USp}}}{}
  \@ifundefined{grpSO}{\newcommand{\grpSO}{\mathrm{SO}}}{}
  \@ifundefined{grpO}{\newcommand{\grpO}{\mathrm{O}}}{}
  \@ifundefined{algOSp}{\newcommand{\algOSp}{\mathfrak{osp}}}{}
  \@ifundefined{algPSU}{\newcommand{\algPSU}{\mathfrak{psu}}}{}
  \@ifundefined{algSU}{\newcommand{\algSU}{\mathfrak{su}}}{}
  \@ifundefined{algSp}{\newcommand{\algSp}{\mathfrak{sp}}}{}
  \@ifundefined{algSL}{\newcommand{\algSL}{\mathfrak{sl}}}{}
  \@ifundefined{algGL}{\newcommand{\algGL}{\mathfrak{gl}}}{}
  \@ifundefined{algU}{\newcommand{\algU}{\mathfrak{u}}}{}
  \@ifundefined{algSO}{\newcommand{\algSO}{\mathfrak{so}}}{}
  \@ifundefined{algO}{\newcommand{\algO}{\mathfrak{o}}}{}
  \@ifundefined{Integers}{\newcommand{\Integers}{\mathbb{Z}}}{}
  \@ifundefined{Reals}{\newcommand{\Reals}{\mathbb{R}}}{} \makeatother

\providecommand{\href}[2]{#2}\begingroup\raggedright\begin{thebibliography}{10}

\bibitem{Aharony:1999ti}
O.~Aharony, S.~S. Gubser, J.~M. Maldacena, H.~Ooguri and Y.~Oz, \emph{Large {N}
  field theories, string theory and gravity},
  \href{https://doi.org/10.1016/S0370-1573(99)00083-6}{\emph{Phys. Rept.}
  {\bfseries 323} (2000) 183}
  [\href{https://arxiv.org/abs/hep-th/9905111}{{\ttfamily hep-th/9905111}}].

\bibitem{Berenstein:2002jq}
D.~E. Berenstein, J.~M. Maldacena and H.~S. Nastase, \emph{Strings in flat
  space and {pp} waves from {$\superN = 4$} super {Y}ang {M}ills}, {\emph{JHEP}
  {\bfseries 0204} (2002) 013}
  [\href{https://arxiv.org/abs/hep-th/0202021}{{\ttfamily hep-th/0202021}}].

\bibitem{Minahan:2002ve}
J.~A. Minahan and K.~Zarembo, \emph{The {B}ethe-ansatz for {$\superN = 4$}
  super {Y}ang-{M}ills}, {\emph{JHEP} {\bfseries 0303} (2003) 013}
  [\href{https://arxiv.org/abs/hep-th/0212208}{{\ttfamily hep-th/0212208}}].

\bibitem{Arutyunov:2009ga}
G.~Arutyunov and S.~Frolov, \emph{Foundations of the {$\AdS{5} \times
  \Sphere^5$} superstring. part {I}},
  \href{https://doi.org/10.1088/1751-8113/42/25/254003}{\emph{J. Phys. A}
  {\bfseries A42} (2009) 254003}
  [\href{https://arxiv.org/abs/0901.4937}{{\ttfamily 0901.4937}}].

\bibitem{Beisert:2010jr}
N.~Beisert, C.~Ahn, L.~F. Alday, Z.~Bajnok, J.~M. Drummond, L.~Freyhult et~al.,
  \emph{{Review of AdS/CFT Integrability: An Overview}},
  \href{https://doi.org/10.1007/s11005-011-0529-2}{\emph{Lett. Math. Phys.}
  {\bfseries 99} (2012) 3} [\href{https://arxiv.org/abs/1012.3982}{{\ttfamily
  1012.3982}}].

\bibitem{Beisert:2005fw}
N.~Beisert and M.~Staudacher, \emph{Long-range {$\grpPSU(2,2|4)$} {B}ethe
  ansaetze for gauge theory and strings},
  \href{https://doi.org/10.1016/j.nuclphysb.2005.06.038}{\emph{Nucl. Phys.}
  {\bfseries B727} (2005) 1}
  [\href{https://arxiv.org/abs/hep-th/0504190}{{\ttfamily hep-th/0504190}}].

\bibitem{Beisert:2005tm}
N.~Beisert, \emph{The {$\algSU(2|2)$} dynamic {$S$}-matrix}, {\emph{Adv. Theor.
  Math. Phys.} {\bfseries 12} (2008) 945}
  [\href{https://arxiv.org/abs/hep-th/0511082}{{\ttfamily hep-th/0511082}}].

\bibitem{Beisert:2006ez}
N.~Beisert, B.~Eden and M.~Staudacher, \emph{Transcendentality and crossing},
  \href{https://doi.org/10.1088/1742-5468/2007/01/P01021}{\emph{J. Stat. Mech.}
  {\bfseries 0701} (2007) P01021}
  [\href{https://arxiv.org/abs/hep-th/0610251}{{\ttfamily hep-th/0610251}}].

\bibitem{Ambjorn:2005wa}
J.~Ambj{\o}rn, R.~A. Janik and C.~Kristjansen, \emph{Wrapping interactions and
  a new source of corrections to the spin-chain/string duality},
  \href{https://doi.org/10.1016/j.nuclphysb.2005.12.007}{\emph{Nucl. Phys.}
  {\bfseries B736} (2006) 288}
  [\href{https://arxiv.org/abs/hep-th/0510171}{{\ttfamily hep-th/0510171}}].

\bibitem{Zamolodchikov:1989cf}
A.~B. Zamolodchikov, \emph{Thermodynamic {B}ethe ansatz in relativistic models.
  {S}caling three state {P}otts and {L}ee-{Y}ang models},
  \href{https://doi.org/10.1016/0550-3213(90)90333-9}{\emph{Nucl. Phys.}
  {\bfseries B342} (1990) 695}.

\bibitem{Arutyunov:2007tc}
G.~Arutyunov and S.~Frolov, \emph{On string {S}-matrix, bound states and
  {TBA}}, \href{https://doi.org/10.1088/1126-6708/2007/12/024}{\emph{JHEP}
  {\bfseries 0712} (2007) 024}
  [\href{https://arxiv.org/abs/0710.1568}{{\ttfamily 0710.1568}}].

\bibitem{Janik:2006dc}
R.~A. Janik, \emph{The {$\AdS{5} \times \Sphere^5$} superstring worldsheet
  {S}-matrix and crossing symmetry},
  \href{https://doi.org/10.1103/PhysRevD.73.086006}{\emph{Phys. Rev.}
  {\bfseries D73} (2006) 086006}
  [\href{https://arxiv.org/abs/hep-th/0603038}{{\ttfamily hep-th/0603038}}].

\bibitem{Arutyunov:2009kf}
G.~Arutyunov and S.~Frolov, \emph{The dressing factor and crossing equations},
  \href{https://doi.org/10.1088/1751-8113/42/42/425401}{\emph{J. Phys.}
  {\bfseries A42} (2009) 425401}
  [\href{https://arxiv.org/abs/0904.4575}{{\ttfamily 0904.4575}}].

\bibitem{Arutyunov:2009mi}
G.~Arutyunov, M.~de~Leeuw and A.~Torrielli, \emph{The bound state {S}-matrix
  for {$\AdS{5} \times \Sphere^5$} superstring},
  \href{https://doi.org/10.1016/j.nuclphysb.2009.03.024}{\emph{Nucl. Phys.}
  {\bfseries B819} (2009) 319}
  [\href{https://arxiv.org/abs/0902.0183}{{\ttfamily 0902.0183}}].

\bibitem{Basso:2015zoa}
B.~Basso, S.~Komatsu and P.~Vieira, \emph{Structure constants and integrable
  bootstrap in planar {$\superN = 4$} {SYM} theory},
  \href{https://arxiv.org/abs/1505.06745}{{\ttfamily 1505.06745}}.

\bibitem{Luscher:1985dn}
M.~L{\"u}scher, \emph{Volume dependence of the energy spectrum in massive
  quantum field theories. 1. {S}table particle states},
  \href{https://doi.org/10.1007/BF01211589}{\emph{Commun. Math. Phys.}
  {\bfseries 104} (1986) 177}.

\bibitem{Luscher:1986pf}
M.~L{\"u}scher, \emph{Volume dependence of the energy spectrum in massive
  quantum field theories. 2. {S}cattering states},
  \href{https://doi.org/10.1007/BF01211097}{\emph{Commun. Math. Phys.}
  {\bfseries 105} (1986) 153}.

\bibitem{Eden:2015ija}
B.~Eden and A.~Sfondrini, \emph{{Three-point functions in ${\cal N}=4$ SYM: the
  hexagon proposal at three loops}},
  \href{https://doi.org/10.1007/JHEP02(2016)165}{\emph{JHEP} {\bfseries 02}
  (2016) 165} [\href{https://arxiv.org/abs/1510.01242}{{\ttfamily
  1510.01242}}].

\bibitem{Basso:2015eqa}
B.~Basso, V.~Goncalves, S.~Komatsu and P.~Vieira, \emph{{Gluing Hexagons at
  Three Loops}},
  \href{https://doi.org/10.1016/j.nuclphysb.2016.04.020}{\emph{Nucl. Phys.}
  {\bfseries B907} (2016) 695}
  [\href{https://arxiv.org/abs/1510.01683}{{\ttfamily 1510.01683}}].

\bibitem{Basso:2017muf}
B.~Basso, V.~Goncalves and S.~Komatsu, \emph{{Structure constants at wrapping
  order}},  \href{https://arxiv.org/abs/1702.02154}{{\ttfamily 1702.02154}}.

\bibitem{Eden:2016xvg}
B.~Eden and A.~Sfondrini, \emph{{Tessellating cushions: four-point functions in
  $\mathcal{N} $ = 4 SYM}},
  \href{https://doi.org/10.1007/JHEP10(2017)098}{\emph{JHEP} {\bfseries 10}
  (2017) 098} [\href{https://arxiv.org/abs/1611.05436}{{\ttfamily
  1611.05436}}].

\bibitem{Fleury:2016ykk}
T.~Fleury and S.~Komatsu, \emph{{Hexagonalization of Correlation Functions}},
  \href{https://doi.org/10.1007/JHEP01(2017)130}{\emph{JHEP} {\bfseries 01}
  (2017) 130} [\href{https://arxiv.org/abs/1611.05577}{{\ttfamily
  1611.05577}}].

\bibitem{Fleury:2017eph}
T.~Fleury and S.~Komatsu, \emph{{Hexagonalization of Correlation Functions II:
  Two-Particle Contributions}},
  \href{https://doi.org/10.1007/JHEP02(2018)177}{\emph{JHEP} {\bfseries 02}
  (2018) 177} [\href{https://arxiv.org/abs/1711.05327}{{\ttfamily
  1711.05327}}].

\bibitem{Drukker:2008pi}
N.~Drukker and J.~Plefka, \emph{{The Structure of n-point functions of chiral
  primary operators in N=4 super Yang-Mills at one-loop}},
  \href{https://doi.org/10.1088/1126-6708/2009/04/001}{\emph{JHEP} {\bfseries
  04} (2009) 001} [\href{https://arxiv.org/abs/0812.3341}{{\ttfamily
  0812.3341}}].

\bibitem{Arutyunov:2009iq}
G.~Arutyunov, M.~de~Leeuw, R.~Suzuki and A.~Torrielli, \emph{{Bound State
  Transfer Matrix for AdS(5) x S**5 Superstring}},
  \href{https://doi.org/10.1088/1126-6708/2009/10/025}{\emph{JHEP} {\bfseries
  10} (2009) 025} [\href{https://arxiv.org/abs/0906.4783}{{\ttfamily
  0906.4783}}].

\bibitem{Eden:2018vug}
B.~Eden, Y.~Jiang, M.~de~Leeuw, T.~Meier, D.~le~Plat and A.~Sfondrini,
  \emph{{Positivity of hexagon perturbation theory}},
  \href{https://arxiv.org/abs/1806.06051}{{\ttfamily 1806.06051}}.

\bibitem{deLeeuw:2019tdd}
M.~de~Leeuw, B.~Eden, D.~l. Plat and T.~Meier, \emph{{Polylogarithms from the
  bound state S-matrix}},  \href{https://arxiv.org/abs/1907.07014}{{\ttfamily
  1907.07014}}.

\bibitem{Goncharov:1998kja}
A.~B. Goncharov, \emph{{Multiple polylogarithms, cyclotomy and modular
  complexes}}, \href{https://doi.org/10.4310/MRL.1998.v5.n4.a7}{\emph{Math.
  Res. Lett.} {\bfseries 5} (1998) 497}
  [\href{https://arxiv.org/abs/1105.2076}{{\ttfamily 1105.2076}}].

\bibitem{Goncharov:2010jf}
A.~B. Goncharov, M.~Spradlin, C.~Vergu and A.~Volovich, \emph{{Classical
  Polylogarithms for Amplitudes and Wilson Loops}},
  \href{https://doi.org/10.1103/PhysRevLett.105.151605}{\emph{Phys. Rev. Lett.}
  {\bfseries 105} (2010) 151605}
  [\href{https://arxiv.org/abs/1006.5703}{{\ttfamily 1006.5703}}].

\bibitem{Bern:2005iz}
Z.~Bern, L.~J. Dixon and V.~A. Smirnov, \emph{{Iteration of planar amplitudes
  in maximally supersymmetric Yang-Mills theory at three loops and beyond}},
  \href{https://doi.org/10.1103/PhysRevD.72.085001}{\emph{Phys. Rev.}
  {\bfseries D72} (2005) 085001}
  [\href{https://arxiv.org/abs/hep-th/0505205}{{\ttfamily hep-th/0505205}}].

\bibitem{Drummond:2007bm}
J.~M. Drummond, J.~Henn, G.~P. Korchemsky and E.~Sokatchev, \emph{{The hexagon
  Wilson loop and the BDS ansatz for the six-gluon amplitude}},
  \href{https://doi.org/10.1016/j.physletb.2008.03.032}{\emph{Phys. Lett.}
  {\bfseries B662} (2008) 456}
  [\href{https://arxiv.org/abs/0712.4138}{{\ttfamily 0712.4138}}].

\bibitem{Golden:2013xva}
J.~Golden, A.~B. Goncharov, M.~Spradlin, C.~Vergu and A.~Volovich,
  \emph{{Motivic Amplitudes and Cluster Coordinates}},
  \href{https://doi.org/10.1007/JHEP01(2014)091}{\emph{JHEP} {\bfseries 01}
  (2014) 091} [\href{https://arxiv.org/abs/1305.1617}{{\ttfamily 1305.1617}}].

\bibitem{Dixon:2011pw}
L.~J. Dixon, J.~M. Drummond and J.~M. Henn, \emph{{Bootstrapping the three-loop
  hexagon}}, \href{https://doi.org/10.1007/JHEP11(2011)023}{\emph{JHEP}
  {\bfseries 11} (2011) 023} [\href{https://arxiv.org/abs/1108.4461}{{\ttfamily
  1108.4461}}].

\bibitem{Dixon:2013eka}
L.~J. Dixon, J.~M. Drummond, M.~von Hippel and J.~Pennington, \emph{{Hexagon
  functions and the three-loop remainder function}},
  \href{https://doi.org/10.1007/JHEP12(2013)049}{\emph{JHEP} {\bfseries 12}
  (2013) 049} [\href{https://arxiv.org/abs/1308.2276}{{\ttfamily 1308.2276}}].

\bibitem{Dixon:2014voa}
L.~J. Dixon, J.~M. Drummond, C.~Duhr and J.~Pennington, \emph{{The four-loop
  remainder function and multi-Regge behavior at NNLLA in planar N = 4
  super-Yang-Mills theory}},
  \href{https://doi.org/10.1007/JHEP06(2014)116}{\emph{JHEP} {\bfseries 06}
  (2014) 116} [\href{https://arxiv.org/abs/1402.3300}{{\ttfamily 1402.3300}}].

\bibitem{Caron-Huot:2016owq}
S.~Caron-Huot, L.~J. Dixon, A.~McLeod and M.~von Hippel, \emph{{Bootstrapping a
  Five-Loop Amplitude Using Steinmann Relations}},
  \href{https://doi.org/10.1103/PhysRevLett.117.241601}{\emph{Phys. Rev. Lett.}
  {\bfseries 117} (2016) 241601}
  [\href{https://arxiv.org/abs/1609.00669}{{\ttfamily 1609.00669}}].

\bibitem{Golden:2014xqa}
J.~Golden, M.~F. Paulos, M.~Spradlin and A.~Volovich, \emph{{Cluster
  Polylogarithms for Scattering Amplitudes}},
  \href{https://doi.org/10.1088/1751-8113/47/47/474005}{\emph{J. Phys.}
  {\bfseries A47} (2014) 474005}
  [\href{https://arxiv.org/abs/1401.6446}{{\ttfamily 1401.6446}}].

\bibitem{Dixon:2016nkn}
L.~J. Dixon, J.~Drummond, T.~Harrington, A.~J. McLeod, G.~Papathanasiou and
  M.~Spradlin, \emph{{Heptagons from the Steinmann Cluster Bootstrap}},
  \href{https://doi.org/10.1007/JHEP02(2017)137}{\emph{JHEP} {\bfseries 02}
  (2017) 137} [\href{https://arxiv.org/abs/1612.08976}{{\ttfamily
  1612.08976}}].

\bibitem{Eden:2000qp}
B.~U. Eden, P.~S. Howe, A.~Pickering, E.~Sokatchev and P.~C. West, \emph{{Four
  point functions in N=2 superconformal field theories}},
  \href{https://doi.org/10.1016/S0550-3213(00)00218-2}{\emph{Nucl. Phys.}
  {\bfseries B581} (2000) 523}
  [\href{https://arxiv.org/abs/hep-th/0001138}{{\ttfamily hep-th/0001138}}].

\bibitem{DHoker:1998vkc}
E.~D'Hoker, D.~Z. Freedman and W.~Skiba, \emph{{Field theory tests for
  correlators in the AdS / CFT correspondence}},
  \href{https://doi.org/10.1103/PhysRevD.59.045008}{\emph{Phys. Rev.}
  {\bfseries D59} (1999) 045008}
  [\href{https://arxiv.org/abs/hep-th/9807098}{{\ttfamily hep-th/9807098}}].

\bibitem{Eden:1999gh}
B.~Eden, P.~S. Howe and P.~C. West, \emph{{Nilpotent invariants in N=4 SYM}},
  \href{https://doi.org/10.1016/S0370-2693(99)00705-4}{\emph{Phys. Lett.}
  {\bfseries B463} (1999) 19}
  [\href{https://arxiv.org/abs/hep-th/9905085}{{\ttfamily hep-th/9905085}}].

\bibitem{Eden:2017ozn}
B.~Eden, Y.~Jiang, D.~le~Plat and A.~Sfondrini, \emph{{Colour-dressed hexagon
  tessellations for correlation functions and non-planar corrections}},
  \href{https://doi.org/10.1007/JHEP02(2018)170}{\emph{JHEP} {\bfseries 02}
  (2018) 170} [\href{https://arxiv.org/abs/1710.10212}{{\ttfamily
  1710.10212}}].

\bibitem{Howe:1999hz}
P.~S. Howe, C.~Schubert, E.~Sokatchev and P.~C. West, \emph{{Explicit
  construction of nilpotent covariants in N=4 SYM}},
  \href{https://doi.org/10.1016/S0550-3213(99)00768-3}{\emph{Nucl. Phys.}
  {\bfseries B571} (2000) 71}
  [\href{https://arxiv.org/abs/hep-th/9910011}{{\ttfamily hep-th/9910011}}].

\bibitem{Alday:2010zy}
L.~F. Alday, B.~Eden, G.~P. Korchemsky, J.~Maldacena and E.~Sokatchev,
  \emph{{From correlation functions to Wilson loops}},
  \href{https://doi.org/10.1007/JHEP09(2011)123}{\emph{JHEP} {\bfseries 09}
  (2011) 123} [\href{https://arxiv.org/abs/1007.3243}{{\ttfamily 1007.3243}}].

\bibitem{Arutyunov:2006yd}
G.~Arutyunov, S.~Frolov and M.~Zamaklar, \emph{The {Z}amolodchikov-{F}addeev
  algebra for {$\AdS{5} \times \Sphere^5$} superstring}, {\emph{JHEP}
  {\bfseries 0704} (2007) 002}
  [\href{https://arxiv.org/abs/hep-th/0612229}{{\ttfamily hep-th/0612229}}].

\bibitem{Basso:2010in}
B.~Basso, \emph{Exciting the {GKP} string at any coupling},
  \href{https://doi.org/10.1016/j.nuclphysb.2011.12.010}{\emph{Nucl. Phys.}
  {\bfseries B857} (2010) 254}
  [\href{https://arxiv.org/abs/1010.5237}{{\ttfamily 1010.5237}}].

\bibitem{Basso:2013vsa}
B.~Basso, A.~Sever and P.~Vieira, \emph{{Spacetime and Flux Tube S-Matrices at
  Finite Coupling for N=4 Supersymmetric Yang-Mills Theory}},
  \href{https://doi.org/10.1103/PhysRevLett.111.091602}{\emph{Phys. Rev. Lett.}
  {\bfseries 111} (2013) 091602}
  [\href{https://arxiv.org/abs/1303.1396}{{\ttfamily 1303.1396}}].

\bibitem{Basso:2013aha}
B.~Basso, A.~Sever and P.~Vieira, \emph{{Space-time S-matrix and Flux tube
  S-matrix II. Extracting and Matching Data}},
  \href{https://doi.org/10.1007/JHEP01(2014)008}{\emph{JHEP} {\bfseries 01}
  (2014) 008} [\href{https://arxiv.org/abs/1306.2058}{{\ttfamily 1306.2058}}].

\bibitem{Basso:2014koa}
B.~Basso, A.~Sever and P.~Vieira, \emph{{Space-time S-matrix and Flux-tube
  S-matrix III. The two-particle contributions}},
  \href{https://doi.org/10.1007/JHEP08(2014)085}{\emph{JHEP} {\bfseries 08}
  (2014) 085} [\href{https://arxiv.org/abs/1402.3307}{{\ttfamily 1402.3307}}].

\bibitem{Basso:2014nra}
B.~Basso, A.~Sever and P.~Vieira, \emph{{Space-time S-matrix and Flux-tube
  S-matrix IV. Gluons and Fusion}},
  \href{https://doi.org/10.1007/JHEP09(2014)149}{\emph{JHEP} {\bfseries 09}
  (2014) 149} [\href{https://arxiv.org/abs/1407.1736}{{\ttfamily 1407.1736}}].

\bibitem{Cordova:2016woh}
L.~Córdova, \emph{{Hexagon POPE: effective particles and tree level
  resummation}}, \href{https://doi.org/10.1007/JHEP01(2017)051}{\emph{JHEP}
  {\bfseries 01} (2017) 051}
  [\href{https://arxiv.org/abs/1606.00423}{{\ttfamily 1606.00423}}].

\bibitem{Lam:2016rel}
H.~T. Lam and M.~von Hippel, \emph{{Resumming the POPE at One Loop}},
  \href{https://doi.org/10.1007/JHEP12(2016)011}{\emph{JHEP} {\bfseries 12}
  (2016) 011} [\href{https://arxiv.org/abs/1608.08116}{{\ttfamily
  1608.08116}}].

\bibitem{Smirnov:1999gc}
V.~A. Smirnov, \emph{{Analytical result for dimensionally regularized massless
  on shell double box}},
  \href{https://doi.org/10.1016/S0370-2693(99)00777-7}{\emph{Phys. Lett.}
  {\bfseries B460} (1999) 397}
  [\href{https://arxiv.org/abs/hep-ph/9905323}{{\ttfamily hep-ph/9905323}}].

\bibitem{Brown:2009ta}
F.~C.~S. Brown, \emph{{On the periods of some Feynman integrals}}, {\emph{Comm.
  Math. Phys.} {\bfseries 287} (2009) 925}
  [\href{https://arxiv.org/abs/0910.0114}{{\ttfamily 0910.0114}}].

\bibitem{Panzer:2014caa}
E.~Panzer, \emph{{Algorithms for the symbolic integration of hyperlogarithms
  with applications to Feynman integrals}},
  \href{https://doi.org/10.1016/j.cpc.2014.10.019}{\emph{Comput. Phys. Commun.}
  {\bfseries 188} (2015) 148}
  [\href{https://arxiv.org/abs/1403.3385}{{\ttfamily 1403.3385}}].

\bibitem{Panzer:2015ida}
E.~Panzer, \emph{{Feynman integrals and hyperlogarithms}}, Ph.D. thesis,
  Humboldt U., 2015.
\newblock \href{https://arxiv.org/abs/1506.07243}{{\ttfamily 1506.07243}}.
\newblock 10.18452/17157.

\bibitem{Drummond:2013nda}
J.~Drummond, C.~Duhr, B.~Eden, P.~Heslop, J.~Pennington and V.~A. Smirnov,
  \emph{{Leading singularities and off-shell conformal integrals}},
  \href{https://doi.org/10.1007/JHEP08(2013)133}{\emph{JHEP} {\bfseries 08}
  (2013) 133} [\href{https://arxiv.org/abs/1303.6909}{{\ttfamily 1303.6909}}].

\bibitem{Eden:2011we}
B.~Eden, P.~Heslop, G.~P. Korchemsky and E.~Sokatchev, \emph{{Hidden symmetry
  of four-point correlation functions and amplitudes in N=4 SYM}},
  \href{https://doi.org/10.1016/j.nuclphysb.2012.04.007}{\emph{Nucl. Phys.}
  {\bfseries B862} (2012) 193}
  [\href{https://arxiv.org/abs/1108.3557}{{\ttfamily 1108.3557}}].

\bibitem{Chicherin:2015edu}
D.~Chicherin, J.~Drummond, P.~Heslop and E.~Sokatchev, \emph{{All three-loop
  four-point correlators of half-BPS operators in planar $ \mathcal{N} $ = 4
  SYM}}, \href{https://doi.org/10.1007/JHEP08(2016)053}{\emph{JHEP} {\bfseries
  08} (2016) 053} [\href{https://arxiv.org/abs/1512.02926}{{\ttfamily
  1512.02926}}].

\bibitem{Brown:2004Logs}
F.~C.~S. Brown, \emph{{}}, {\emph{C. R. Acad. Sci. Paris} {\bfseries I} (2004)
  338}.

\bibitem{Broedel:2015nfp}
J.~Broedel and M.~Sprenger, \emph{{Six-point remainder function in
  multi-Regge-kinematics: an efficient approach in momentum space}},
  \href{https://doi.org/10.1007/JHEP05(2016)055}{\emph{JHEP} {\bfseries 05}
  (2016) 055} [\href{https://arxiv.org/abs/1512.04963}{{\ttfamily
  1512.04963}}].

\bibitem{Huber:2005yg}
T.~Huber and D.~Maitre, \emph{{HypExp: A Mathematica package for expanding
  hypergeometric functions around integer-valued parameters}},
  \href{https://doi.org/10.1016/j.cpc.2006.01.007}{\emph{Comput. Phys. Commun.}
  {\bfseries 175} (2006) 122}
  [\href{https://arxiv.org/abs/hep-ph/0507094}{{\ttfamily hep-ph/0507094}}].

\bibitem{Huber:2007dx}
T.~Huber and D.~Maitre, \emph{{HypExp 2, Expanding Hypergeometric Functions
  about Half-Integer Parameters}},
  \href{https://doi.org/10.1016/j.cpc.2007.12.008}{\emph{Comput. Phys. Commun.}
  {\bfseries 178} (2008) 755}
  [\href{https://arxiv.org/abs/0708.2443}{{\ttfamily 0708.2443}}].

\bibitem{Drukker:2009sf}
N.~Drukker and J.~Plefka, \emph{Superprotected n-point correlation functions of
  local operators in {$\superN = 4$} super {Y}ang-{M}ills},
  \href{https://doi.org/10.1088/1126-6708/2009/04/052}{\emph{JHEP} {\bfseries
  04} (2009) 052} [\href{https://arxiv.org/abs/0901.3653}{{\ttfamily
  0901.3653}}].

\end{thebibliography}\endgroup

\end{document}